\documentclass[12pt,preprint]{aastex}
\usepackage{natbib}
\newcommand{\MSun}{M_{\odot}}
\newcommand{\Msun}{M_{\odot}}
\newcommand{\Msunkpc}{\; \MSun~{\rm kpc}^{-2}}
\newcommand{\MSunkpc}{\; \MSun~{\rm kpc}^{-2}}

\slugcomment{April 2007 draft}

\author{
Timothy A. Reichard\altaffilmark{1}, Timothy
M. Heckman\altaffilmark{1}, Gregory Rudnick\altaffilmark{2,5}, Jarle
Brinchmann\altaffilmark{3}, Guinevere Kauffmann\altaffilmark{4}}

\altaffiltext{1}{Department of Physics and Astronomy, The Johns Hopkins University, 3400 North Charles Street, Baltimore, MD 21218-2686.}
\altaffiltext{2}{National Optical Astronomy Observatory (NOAO), 950 N. Cherry Ave., Tucson, AZ 85719, USA}
\altaffiltext{3}{Centro de Astrofisica da Universidade do Porto, Rua das Estrelas, 4150-762 Porto, Portugal.}
\altaffiltext{4}{Max-Planck-Institut f\"{u}r Astrophysik, D-85748 Garching, Germany.}
\altaffiltext{5}{Goldberg Fellow}
\begin{document}

\title{The Lopsidedness of Present-Day Galaxies: Results from the Sloan Digital Sky Survey}

\begin{abstract}
Large-scale asymmetries in the stellar mass distribution in galaxies
are believed to trace non-equilibrium situations in the luminous
and/or dark matter component. These may arise in the aftermath of
events like mergers, accretion, and tidal interactions. These events
are key in the evolution of galaxies. In this paper we quantify the
large-scale lopsidedness of light distributions in 25155 galaxies at
$z < 0.06$ from the Sloan Digital Sky Survey Data Release 4 using the
$m = 1$ azimuthal Fourier mode.  We show that the lopsided
distribution of light is primarily due to a corresponding lopsidedness
in the stellar mass distribution. Observational effects, such as
seeing, Poisson noise, and inclination, introduce only small errors in
lopsidedness for the majority of this sample. We find that
lopsidedness correlates strongly with other basic galaxy structural
parameters: galaxies with low concentration, stellar mass, and stellar
surface mass density tend to be lopsided, while galaxies with high
concentration, mass, and density are not. We find that the strongest
and most fundamental relationship between lopsidedness and the other
structural parameters is with the surface mass density. We also find,
in agreement with previous studies, that lopsidedness tends to
increase with radius. Both these results may be understood as a
consequence of several factors. The outer regions of galaxies and
low-density galaxies are more susceptible to tidal perturbations, and
they also have longer dynamical times (so lopsidedness will last
longer). They are also more likely to be affected by any underlying
asymmetries in the dark matter halo.

\end{abstract}

\section{Introduction}

%\citet{rz95}: 18 galaxies. Galaxy disks exhibit a wide variety of shapes visible in the near-IR.  They are due to distortions in surface density distributions, not mass-to-light ratios.  Stellar velocities in nonaxisymmetric galaxies differ by 3-6\% from those in symmetric ones.

%\citet{zr97}: 60 galaxies, magnitude-limited.  30\% of galaxies have $A_1 > 0.2$.  Lopsided mass distributions remain long enough to indicate past interactions, not just ongoing ones.  Often the companion is not obvious, either merged or fled.

%\citet{rr98}: 54 early-type disk galaxies.  Low SFRs to reduce possibility of asymmetric stellar distributions and increase possibilty of light->mass asymmetry tracing.  Traces mass dependence because VRI bands show similar lopsidedness and it's OK to go shorter than I and K bands.  20\% of galaxies have lopsidedness above 0.19.  

%\citet{rrk00}: Correlations between lopsidedness and recent SF, and current SF were found, lopsided = star-forming.  Significant fractions of stellar content can be created in a short time from minor mergers and on a similar timescale (1 Gyr). 

It has long been recognized that galaxies show large-scale asymmetries
in their structure \citep{bl+80}. Lopsided galaxies have such
asymmetries where one side of their disk is more massive and/or more
extended than the opposite side.  This "lopsidedness" can be traced in
the spatial structure of the stars \citep{rz95} and/or the HI gas
\citep{rs94} and/or in the large-scale kinematics of this material
\citep{ss+99}.

There are a variety of mechanisms or events that have been proposed to
produce the observed lopsidedness. All of them involve a
time-dependent non-equilibrium dynamical state, in most cases
triggered through an external process. Such external processes are a
natural consequence of the standard Lambda Cold Dark Matter
cosmological framework. This implies that galaxies assemble
hierarchically (a process that is on-going). Examples that can lead to
lopsidedness include a minor merger (\citealt{wm+96};
\citealt{zr97}), the tidal interaction resulting from a close
encounter between roughly equal-mass galaxies \citep{kl+02},
and the asymmetric accretion of intergalactic gas into the disk
(\citealt{bc+05}; \citealt{kk+05}). Other mechanisms involve the
dark matter halo: stars and gas orbiting in a lopsided dark matter
halo (\citealt{w94}; \citealt{j97}; \citealt{j99}) or a stellar/gas
disk that is offset with respect to the center of the dark matter halo
(\citealt{ls98}; \citealt{ns+01}). These
also involve past tidal interactions and/or mergers that have
perturbed the dark matter halo, but such perturbations may be quite
long-lived. Finally, dynamical processes internal to the disk that
lead to mildly lopsided distributions have also been investigated (\citealt{st+90}; \citealt{st96}; \citealt{mt97}).

A variety of programs to study lopsidedness have been undertaken over
the past decade. Most of these investigations have studied the
lopsided distribution of the stellar component through analysis of
optical and near-infrared images. \citet{zr97} studied a
magnitude-limited sample of 60 field spiral galaxies. They measured
lopsidedness as the radially averaged, azimuthal $m=1$ Fourier
amplitude $A_1$ of the light (see Section \ref{sec:error} below) and
computed lopsidedness between 1.5 and 2.5 scale lengths in the
galactic disks. The value of $A_1$ indicates the typical large-scale
variation in mass density from side to opposite side at the same
distance from the galactic center. The mass density typically varies
from between $1 \pm A_1$ times the average density at the same radius.
They found that $\sim 30$\% of field spiral galaxies exhibited
significant lopsidedness ($A_1 > 0.2$).  \citet{rr98} followed up this
work by studying lopsidedness in 54 early-type galaxies and found that
$\sim 20$\% had $A_1 > 0.19$.
\citet{cbj00} studied a sample of 113 $z < 0.01$ galaxies (elliptical, spiral,
and irregular) but used a 180-degree rotational asymmetry measure
$A_{180}$.  They found that asymmetry is strongly dependent on
morphological type, with lower asymmetry in elliptical and lenticular
galaxies and higher asymmetry in late-type disk and irregular
galaxies. More recently, \citet{bc+05} have measured the Fourier $A_1$
parameter for 149 galaxies in the Ohio State University Bright Galaxy
Survey. They confirmed that a large fraction of galaxies have
significant lopsidedness in their stellar disks, with late-type
galaxies being more lopsided.

Lopsidedness in the {\it light} distribution can be produced by either
a corresponding asymmetry in the underlying {\it mass} distribution in
the stellar population or by large-scale variations in the
mass-to-light ratio (e.g., from star formation and dust
obscuration). \citet{rz95} investigated this issue with a sample of 18
face-on spiral galaxies imaged in the K$^\prime$ (2.2$\mu$ m) band
where the effect of young stars or dust is minimized. They found that
about a third of the sample showed significant lopsidedness (similar
to results from optical investigations). Similarly, \citet{rr98} found
that lopsidedness in early-type disk galaxies is nearly identical when
observed in the $V$, $R$, and $I$ bands. They concluded that an
asymmetric mass distribution then accounts for the majority of the
asymmetry in the light distribution in these galaxies.

Lopsidedness has also been studied in the distribution of HI
gas. Since the HI can frequently be traced to significantly larger
radii than the stars, these investigations are highly complementary to
the optical image analysis. Due to the time-consuming nature of HI
interferometric mapping, only modest size samples have been analyzed
in this way \citep{ss+99}. On the other hand, \citet{rs94} have examined the global HI line profiles for roughly 1700
galaxies, and shown that at least 50\% are significantly asymmetric
(confirming that the large-scale HI distribution is frequently
lopsided). HI maps also show that – apart from a lopsided distribution
of the gas – the HI rotation curves are often asymmetric \citep{ss+99}. The connection between the phenomena of structural and
kinematic lopsidedness in galaxies is not yet clear \citep{ss+99}.

Despite these diverse investigations and the abundance of proposed
models, the origin of lopsidedness remains unsettled. For models
involving tidal interactions or minor mergers, there is an expected
link between lopsidedness and the local environment. The evidence in
this regard has been mixed (e.g., \citealt{wp04};
\citealt{bc+05}; \citealt{aj+06,aj+07}; \citet{dc+07}).

%\citet{dc+07} have undertaken
%the most comprehensive investigation so far of this issue. Using the
%rotational asymmetry measure $A_{180}$ to study a sample of over 3000
%galaxies, they find that close pairs of galaxies are more asymmetric
%than other galaxies and that the asymmetry increases as the pair
%separation decreases. They conclude that these global asymmetries
%trace recent tidal interactions or mergers.

The investigations summarized above have all involved relatively small
samples of galaxies, making it difficult to assess the overall
distribution of asymmetry or lopsidedness as a function of the basic
parameters that characterize the structure of galaxies. This is the
first of three papers in which we use the wealth of data available
from the Sloan Digital Sky Survey (SDSS) to extend these studies of
small samples (of-order one hundred galaxies) to large samples (tens
of thousands). In this paper, we describe our sample selection and
methodology. We also relate lopsidedness to the basic structural
properties of the galaxies. In Paper II we will investigate the
connection between lopsidedness and both star formation and black hole
growth in galaxies. Finally, in Paper III we will examine the
connection between lopsidedness and the local galaxy environment.

In \S\ref{sec:data}, we begin by presenting an initial low-redshift
sample from the SDSS and describe the observations and properties for
its galaxies.  Next, we explain our lopsidedness calculation. In
\S\ref{sec:syserr}, we address the major data quality issues that
limit the reliability of the measurements for portions of the
sample. On this basis, we apply cuts on the observational parameters
to weed out the problematic cases for our subsequent analysis. Next,
\S\ref{sec:lopprops} describes the lopsidedness of galactic light
distributions in different optical/near-IR bands, its correspondence
with lopsided mass distributions, and its radial dependence.  We then
examine the relationship between lopsidedness and the basic structural
properties of galaxies in \S\ref{sec:lopsfh}.  Finally, we summarize
our findings in \S\ref{sec:summary}.

\section{Data \& Basic Methodology \label{sec:data}}

\subsection{Initial Sample}

The initial sample of galaxies was taken from the Sloan Digital Sky Survey
\citep{y+00,s+02}, a large survey of photometric and spectroscopic
data across $\pi$ sr. of the northern sky. The sample is derived from SDSS Data Release 4 \citep{dr4}. The survey's dedicated 2.5-m telescope \citep{g+06} at Apache Point Observatory uses a unique CCD camera
\citep{g+98} and drift-scanning to obtain $u$-, $g$-, $r$-, $i$-, and
$z$-band photometry \citep{f+06,h+01,i+04,sm+02,t+06}. The pixel scale
is 0\farcs396/px.  Fiber spectroscopy is obtained using
3\arcsec~fibers and results in wavelength coverage between
3800$-$9200~\AA~at a resolution $R=\lambda/\delta\lambda = 1850-2200$.

As we will show below, meaningful measurements of lopsidedness impose
requirements on the angular size and signal-to-noise in the galaxy
image.  These criteria are not met by the full SDSS Main galaxy sample
(median redshift $\sim0.1$ and magnitude $r <$ 17.8; citealt{swl+02}). Accordingly, our
initial galaxy sample was selected from the Main sample with a simple
redshift cut ($z \le 0.06$). It contains 67107 galaxies. The SDSS
photometric pipeline, PHOTO \citep{lgi+01}, provides Petrosian
apparent magnitudes and half- and 90\%-light radii ($R_{50}$ and
$R_{90}$) in each of the five bands, along with seeing conditions and
other photometric and structural properties. We use the methodology
described in \citet{khw+03a} to use the spectral information to derive
$z$-band mass-to-light ratios and hence stellar masses ($M_*$). These
stellar masses and the $z$-band half-light radii are then used to
measure stellar surface mass densities ($\mu_*$, defined as the mean
mass per unit area inside the half-light radius; see
\citealt{khw+03b}).

\subsection{Measuring Lopsidedness \label{sec:error}}

Lopsidedness in galactic light distributions has been computed in
recent years by two useful approaches.  One approach, used by
\citet{rz95} and in later work, is to perform an azimuthal Fourier
decomposition of galaxy light.  The first mode, $A_1$, quantifies the
large-scale overabundance of light in one side of a galaxy with a
corresponding under-abundance in the opposite side.  The first mode
quantifies a galaxy-wide lopsidedness that can be quickly computed for
large numbers of galaxies.  In the second approach, detailed in
\citet{avg+96}, a different asymmetry index $A_{180}$ was devised by
subtracting a 180$^\circ$ rotated image from the original galaxy image
and summing the residual light compared to the total galaxy light.
This lopsidedness measure is sensitive to both large- and small-scale
variations in symmetry.  To convert $A_{180}$ into a measure of only
large-scale lopsidedness, a smoothing filter can be applied beforehand
to the image \citep{c03}.  Our interest is to examine lopsidedness in
tens of thousands of galaxies, and we have determined that the modal
lopsidedness $A_1$ algorithm is computationally less expensive than
$A_{180}$.  We will therefore take the radially averaged first mode
strength, $A_1$, as the measure of lopsidedness for the galaxy as a
whole in each band, as has been done in previous studies
\citep{rz95,zr97,rr98,rrk00}. The details of the calculation will fill out this section, but we first summarize the general method.

The galaxy light is binned into radial and azimuthal bins, and then a
finite Fourier transform is applied to the surface brightness $\mu$ in
each radial bin.  The transform for the $k$th radial bin is
\begin{equation}
\mu(r_k, \phi) = \sum\limits_{m = 0}^{m_{max}} b_m (r_k) e^{im(\phi - \phi_m (r_k))}. \label{eqn:decomp}
\end{equation}
The transform gives the mode magnitudes $b_m(r_k)$ and phases
$\phi_m(r_k)$ for each mode $m$. The magnitudes are divided by the
zeroth mode (average surface brightness) and radially averaged to calculate
$A_m$ for that galaxy and band, including the lopsidedness $A_1$.

We compute this average over
the radial range between $R_{50}$ and $R_{90}$. These are the
practical limits: we can not go much inside $R_{50}$ because of the
seeing effects discussed below, and can not go much outside $R_{90}$
because the data become unacceptably noisy. As we will show below, the
average radial variation in $A_1$ beyond is modest, and the mean value
for $A_1$ does not depend significantly on the precise choices of the
inner and outer radii.

The transform also yields values of the higher-order modes. The second
mode $A_2$ represents a combination of effects: ellipticity,
inclination, two-arm spiral structure, and barred structure.  Since we
do not deproject the images to a face-on orientation, the amplitude of
the second mode primarily measures the ellipticity of the
image. Higher-order modes ($m > 2$) may include multi-arm spiral
structure as well as Fourier ``ringing'' of features also present in
lower-order modes.  For example, a galaxy with a high ellipticity will
have a strong second mode with weaker even-order modes also present.
Similarly, a lopsided galaxy will have a strong first mode with weaker
odd-order modes present as well.

To begin our process, the galaxy center was determined. Precision centering is
important because moving the center by more than $0.5$ px can increase
the lopsidedness significantly. The inner region of the galaxy was
smoothed in a 3 px radius to reduce the effects of Poisson noise in
determining the center.
\footnote{\citet{cbj00} have employed a
different centering technique with their $A_{180}$ asymmetry measure.
They found that the best center to use for the $A_{180}$ calculation
was the center that minimizes $A_{180}$.  In our multi-mode
calculation, we have found that the center that minimizes the strength
of one mode does not minimize the strengths of other modes.  Thus the
minimization technique would require a different center for each mode.
Since the modes are calculated simultaneously in the Fourier
transform, minimization is not feasible.}  
We assume that the center point of the galaxy is within 3 px of the
brightest pixel of the inner region of the galaxy. That brightest
pixel is our initial estimate of the center point.  To improve this
estimate, the first moment of light was computed in a 3 $\times$ 3 px
box centered on the brightest pixel.  The first moment is the origin
that minimizes the second moment and gives a central position in
fractional units of a pixel that improves upon the brightest pixel
estimate.
%In low-$S/N$ galaxies, Poisson noise can have an important effect
%in centering and introduces more uncertainty into both the center
%position and the modes.  The effect is strongest in the innermost
%region of the galaxy.  To minimize this effect, we always mask the
%inner 3.0 px (1.2$\arcsec$), even if $R_{50}$ is smaller than this
%radius.

Our centering method was used on each image without reference to the
centers determined for the same object in other bands.  This means
that each object was given a separate center point for each band.
Nonetheless, the agreement between these independent center points is good.
Of the three pairs of distances between centers in the
three bands, the angular distance between the $g$- and $i$-band
centers is the greatest.  The $1\sigma$ variation between these bands
is $0\farcs12^{+0.11}_{-0.07}$ ($0.32^{+0.29}_{-0.17}$ px). Some of
the scatter may be attributable to the offset pixel locations between
images of different bands, i.e., the coordinates of the center of the
central pixel in one band will not be the located in the center of a
pixel in another band.  The rest of the scatter is due to variations
in the central structure between the different bands.  Our tests show
that the mode strengths are significantly increased in a given image
if the center point is moved more than 0.5 px from the best center.
Thus the scatter of center points between the different bands should
have no significant affect on the subsequent lopsidedness computation.

The value of the sky background was calculated using the DAOPHOT
package in the IDL Astronomy Library.  For each galaxy and each band,
the sky-subtracted image was partitioned into a polar grid of
logarithmic radial bins between the Petrosian radii $R_{50}$ and
$R_{90}$ and centered at the determined galaxy center.  We selected
the minimum $R_{50}$ and maximum $R_{90}$ determined in the $g$-,
$r$-, and $i$-band images as the inner and outer radii of the grid,
and the same grid size is used in the lopsidedness computation in all
three bands.  The radial bins were further partitioned into equal
azimuthal bins.  The number of radial bins was allowed to vary with
the size of the galaxies by setting the innermost bin size as close to
but no smaller than 1.0 pixel.  Exactly 12 azimuthal bins were used in
each grid so that the first six Fourier modes ($m = 0, 1, 2, \ldots,
5$) could be determined.  Each pixel that was split by the grid into
more than one bin was divided into a $9
\times 9$ array of sub-pixels, and bilinear interpolation of
neighboring pixels' surface brightness assigned each sub-pixel with a
value of surface brightness. Higher-order interpolation, such as using
bicubic splines, showed no significant improvement and thus was not
worth the computational expense. The sub-pixels were then included in
the appropriate polar bin.

Galaxies whose observed light distributions are contaminated with
light from other galaxies and foreground stars pose a problem for this
Fourier decomposition. The contaminating light results in a set of
strong modes that describe the combination of galaxy and overlapping
star or galaxy rather than the desired galaxy by itself.  Foreground
stars can be masked out with some ease once they are located in
images.  We use the DAOPHOT routines of the IDL Astronomy Library to
locate them and then estimate radii where their light drops to half the
level of the $1\sigma$ sky background noise.  For intervening
galaxies, we queried the SDSS PHOTO catalog for galaxy positions and
Petrosian $R_{90}$.  Then a mask image was created that was the same
size as the galaxy image, but each pixel accepted values of either 0
(pixel is masked) or 1 (pixel is ``good'').  The Fourier decomposition
calculation uses only the ``good'' pixels after the stars and galaxies
are masked in the image.  Any pixels that lay within a star, another
galaxy, or other extraneous light source were ignored.

After masking out the unwanted pixels, the surface brightness was
computed in each bin of the grid.  An azimuthal Fourier transform
(Eq.~\ref{eqn:decomp}) was performed in each radial bin to determine a
magnitude $b_m (r_k)$ and phase $\phi_m (r_k)$ for each mode $m$.  The
transform was formulated as a general linear least squares problem
\citep{p02}.  Errors on the magnitudes and phases were deduced from
the diagonal elements of the covariance matrix.  The phase $\phi_m$ of
the $m$th mode is physically identical to phases $\phi_m + 2\pi K/m$
for integers $K$, so $K$ was chosen for each radial bin to minimize
the radial dependence of the phases.  The magnitudes $b_m (r_k)$
were then divided by the zeroth mode to give relative magnitudes $a_m
(r_k) = b_m (r_k)/b_0 (r_k)$.  These values $a_m (r_k)$ give a profile
of lopsidedness ($m = 1$) and higher modes ($m > 1$), each normalized
to the average brightness at radius $r_k$.

For simplicity, we prefer to use a single value of the Fourier modes
in each band rather than a radial profile. We calculate the average of
$a_m$ at each radius between $R_{50}$ and $R_{90}$, weighted by its
error at each radius, to give a single, global measure $\left <a_m
\right>$ for each galaxy in each band.  The weighting naturally counts
the dimmer, outer reaches of the galaxy less than the brighter, inner
region and discounts any radial bin that is too contaminated with
light from other sources.  At radii where an extraneous light source
has been masked out, the azimuthal Fourier transform can fail.  Such
radial bins are excluded from the weighted average.

Since $\left <a_m \right>$ is a positive definite quantity, random errors would
preferentially overestimate the quantity.  We therefore correct the
average in the manner used in the past (e.g., \citealt{rr98}) and
adopt
\begin{equation}
A_m = \sqrt{\left <a_m \right >^2 - (\delta\left <a_m \right >)^2}
\label{eqn:posdef}
\end{equation}
as the strength of the $m$th mode, where $\delta \left <a_m \right >$
is the random error in the weighted mean mode.  We will discuss
systematic and random errors in $A_1$ in the next section
(\S\ref{sec:syserr}).

\begin{figure}[ht]

\epsscale{1.0}
%\plotone{/home/tar/agn/asymm/imreconst/0338-51694-360.ps}
\plotone{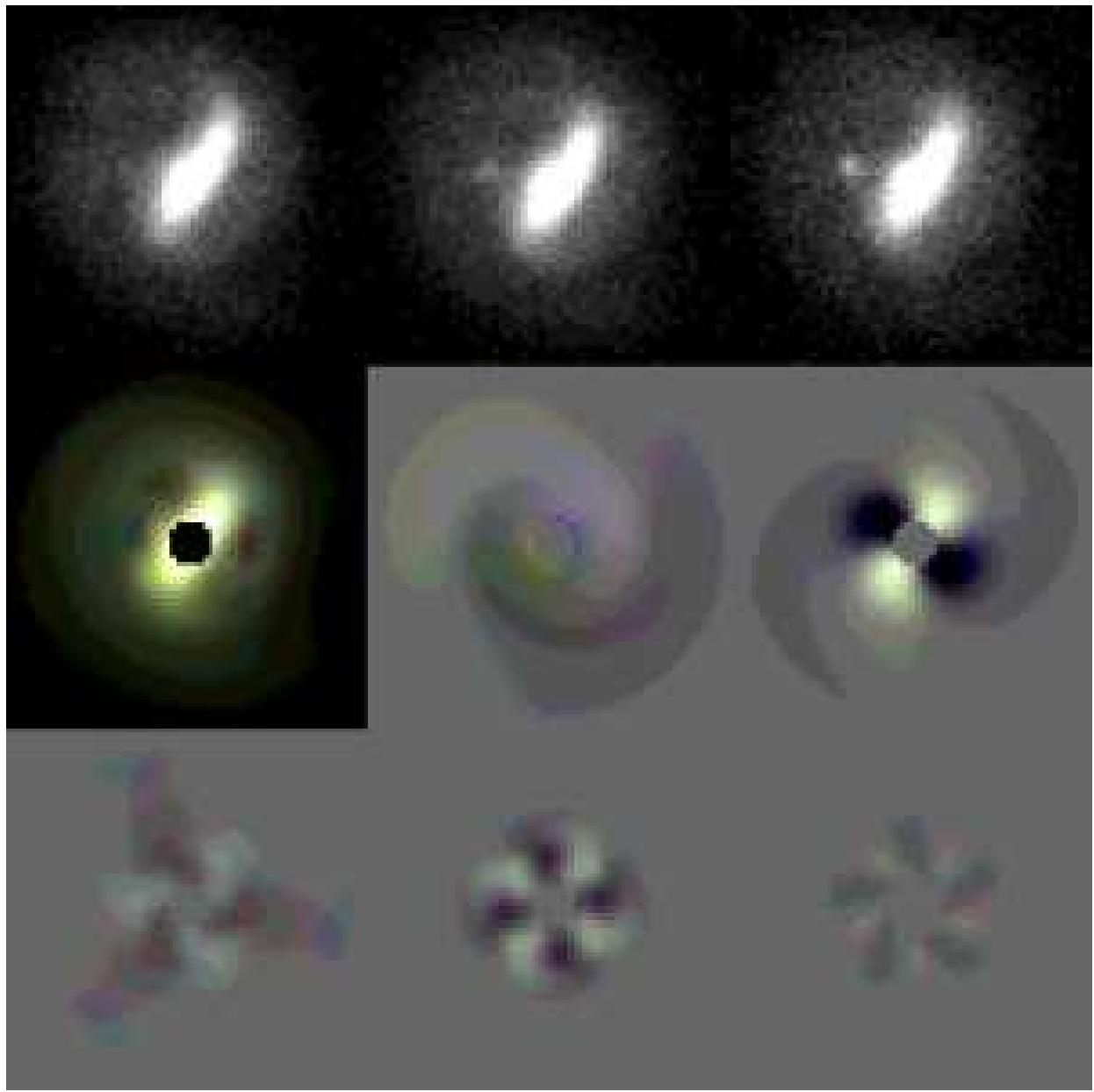}
\caption{Images of the Fourier decomposition of SDSS J125416.38-020204.4 with $A_1^i = 0.29$.  {\em Upper row, left to right:} SDSS $g$-, $r$-, and $i$-band images.  {\em Middle row, left:} Combined $gri$ image reconstructed from Fourier modes.  Light from the $g$-band is colored blue, $r$-band green, and $i$-band red.  {\em Middle row, center and right:} Combined $gri$ images of the first and second modes alone.  {\em Lower row, left to right:} Combined $gri$ images of the third, fourth, and fifth modes alone.}
\label{fig:sampleim}
\end{figure}

As an example, we show the Fourier decomposition of the nearly
face-on, barred spiral galaxy SDSS J125416.38-020204.4 in
Fig.~\ref{fig:sampleim}.  The first three panels show the $g-$, $r-$,
and $i-$band images of this galaxy, a three-color image reconstructed
from Fourier modes, and a three-color image of the $m = 1$ through 5
Fourier modes. The core of the galaxy is omitted in the
three-color images as the decomposition is not attempted in that
region. However, for this illustration, we have extended the Fourier
decomposition to radii interior to $R_{50}$.  The central bar is
described by the strong even-order modes in all three bands at low
radii $2\arcsec < r < 5\arcsec$.  The bar weakens at larger radii, and
the even-order modes then decrease with increasing radii.  At $r >
8\arcsec$, the brightness of the galaxy drops until sky noise becomes
significant.  The lopsidedness in this galaxy is easily seen outside
the bright center of the galaxy.  The bar's brightness falls off more
slowly with radius in the upper-right direction than in the lower-left
direction.  This lopsidedness is described in all three bands by the
significant first mode strength, which rises from near 0.1 to 0.4
between 2$\arcsec$ and 8$\arcsec$.  The radially-averaged
lopsidedness, as computed between $R_{50}$ and $R_{90}$ in the
$i$-band, is 0.29.

\clearpage

\begin{figure}
\centering
\leavevmode
\columnwidth=.30\columnwidth
\begin{minipage}[b]{1.6in}
\includegraphics[width=1.6in]{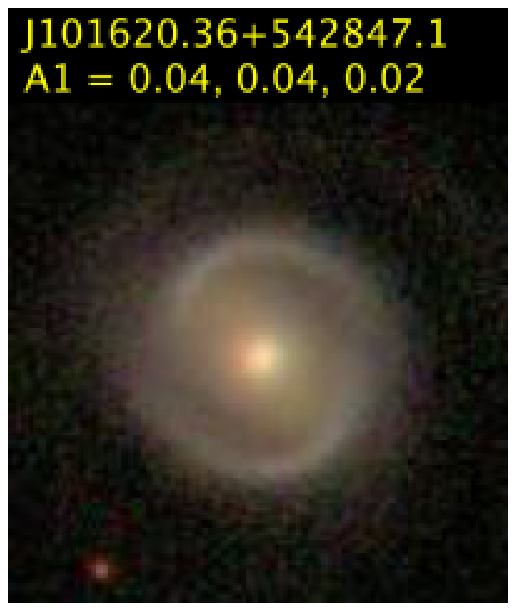} 
%{\scriptsize J101620.36+542847.1
%
%$A_1 = 0.04, 0.04, 0.02$}
\end{minipage}
\hfill
\begin{minipage}[b]{1.6in}
\includegraphics[width=1.6in]{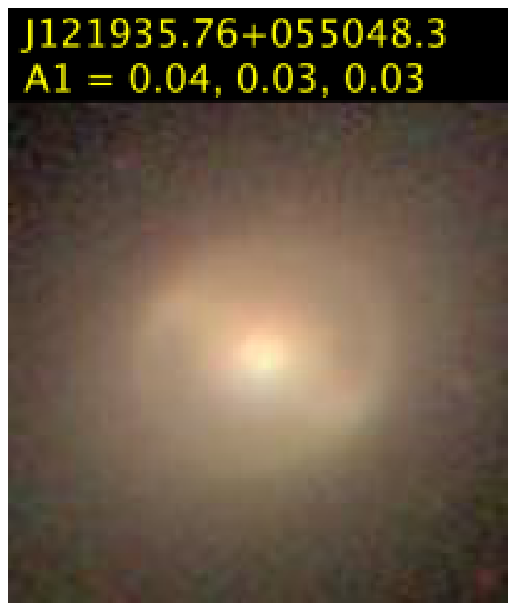} 
%{\scriptsize J121935.76+055048.3
%
%$A_1 = 0.04, 0.03, 0.03$}
\end{minipage}
\hfill
\begin{minipage}[b]{1.6in}
\includegraphics[width=1.6in]{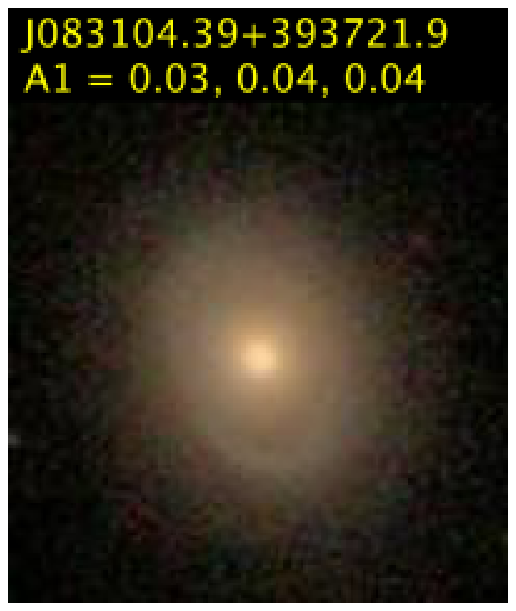}
%{\scriptsize J083104.39+393721.9
%
%$A_1 = 0.03, 0.04, 0.04$}
\end{minipage}
\hfill
\begin{minipage}[b]{1.6in}
\includegraphics[width=1.6in]{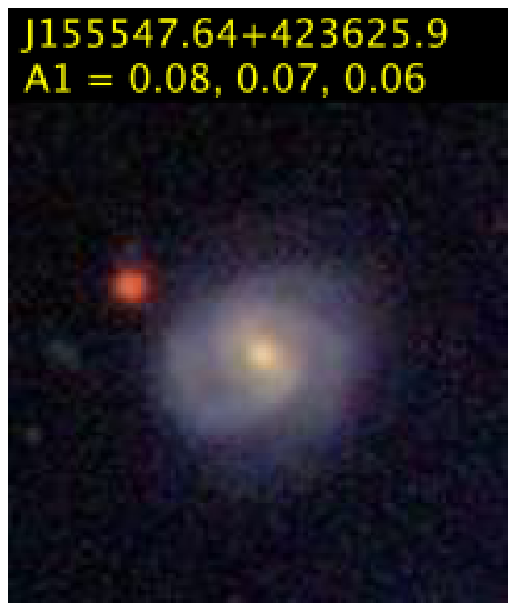}
%{\scriptsize J155547.64+423625.9
%
%$A_1 = 0.08, 0.07, 0.06$}
\end{minipage}
\hfill
\begin{minipage}[b]{1.6in}
\includegraphics[width=1.6in]{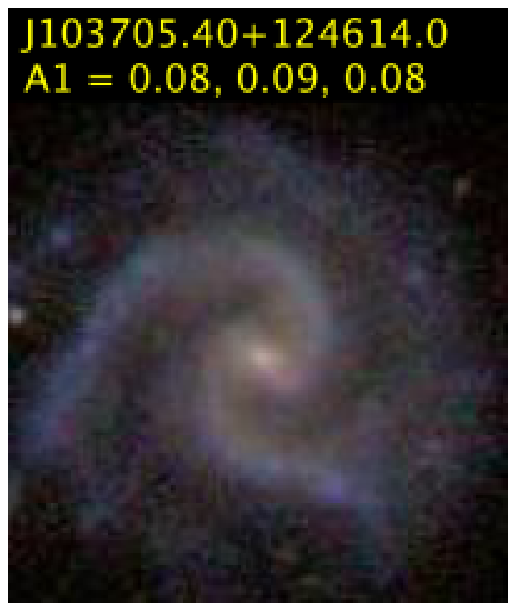}
%{\scriptsize J103705.4+124614
%
%$A_1 = 0.08, 0.09, 0.08$}
\end{minipage}
\hfill
\begin{minipage}[b]{1.6in}
\includegraphics[width=1.6in]{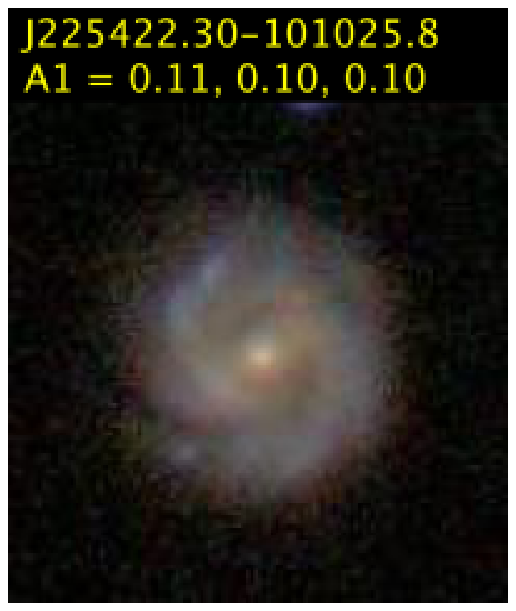}
%{\scriptsize J225422.3-101025.8
%
%$A_1 = 0.11, 0.10, 0.10$}
\end{minipage}
\hfill
\begin{minipage}[b]{1.6in}
\includegraphics[width=1.6in]{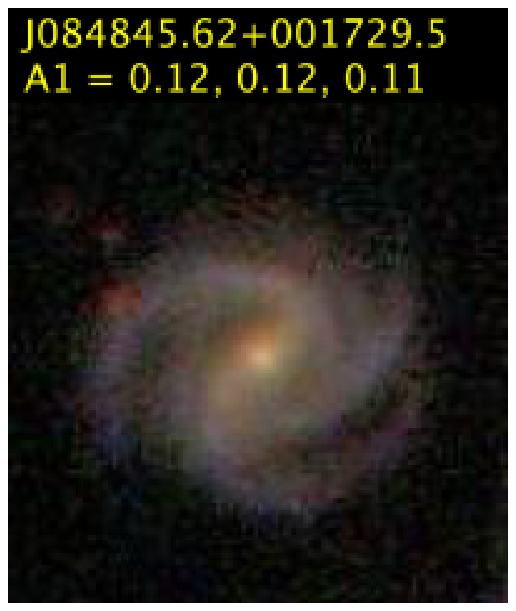}
%{\scriptsize J084845.62+001729.5
%
%$A_1 = 0.12, 0.12, 0.11$}
\end{minipage}
\hfill
\begin{minipage}[b]{1.6in}
\includegraphics[width=1.6in]{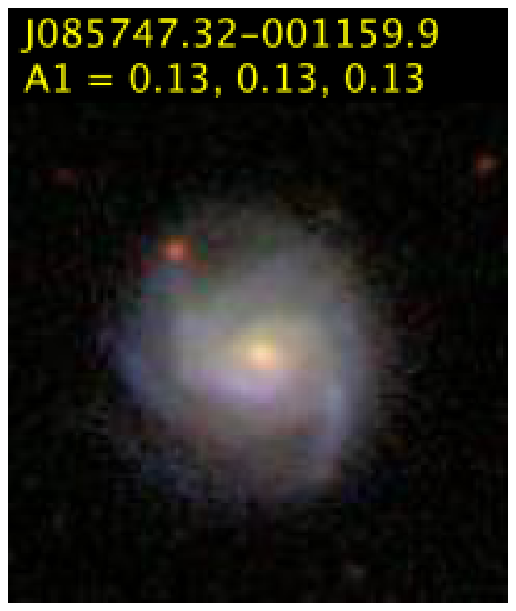}
%{\scriptsize J085747.32-001159.9
%
%$A_1 = 0.13, 0.13, 0.13$}
\end{minipage}
\hfill
\begin{minipage}[b]{1.6in}
\includegraphics[width=1.6in]{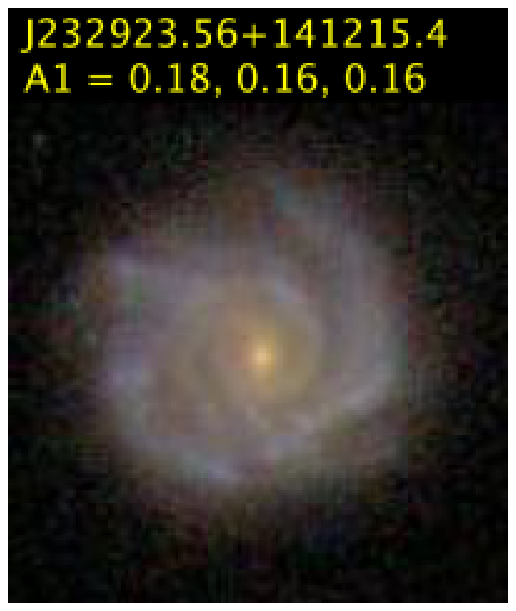}
%{\scriptsize J232923.56+141215.4
%
%$A_1 = 0.18, 0.16, 0.16$}
\end{minipage}
\hfill
\begin{minipage}[b]{1.6in}

\includegraphics[width=1.6in]{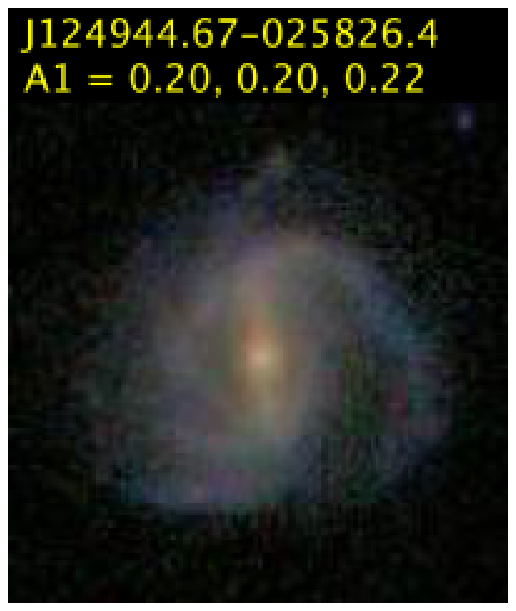}
%{\scriptsize J124944.67-025826.4
%
%$A_1 = 0.20, 0.20, 0.22$}
\end{minipage}
\hfill
\begin{minipage}[b]{1.6in}
\includegraphics[width=1.6in]{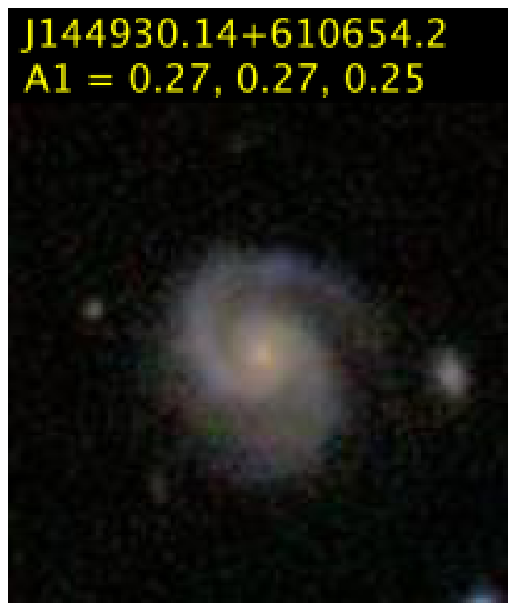}
%{\scriptsize J144930.14+610654.2
%
%$A_1 = 0.27, 0.27, 0.25$}
\end{minipage}
\hfill
\begin{minipage}[b]{1.6in}
\includegraphics[width=1.6in]{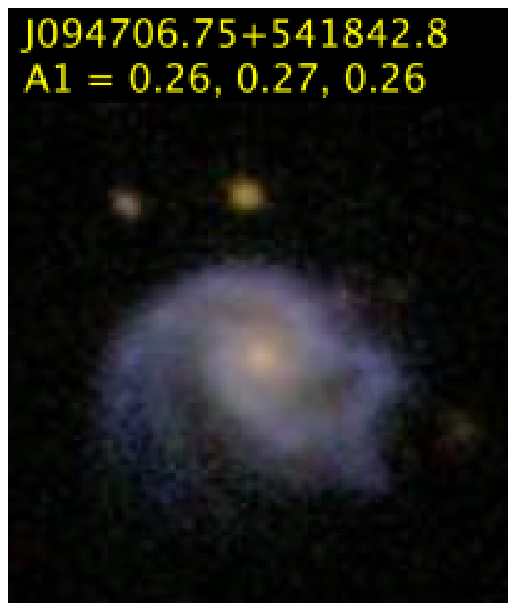}
%{\scriptsize J094706.75+541842.8
%

%$A_1 = 0.26, 0.27, 0.26$}
\end{minipage}
\hfill
\caption{Twelve late-type SDSS galaxies shown with their $g$-, $r$-, and $i$-band lopsidedness.  They are arranged by increasing lopsidedness toward the right in each row, least lopsided in the uppermost row and most lopsided in the lowermost row.  Clear distortions in shape and/or brightness are visible in the galaxies with $A_1 > 0.08$. \label{fig:mosaic}}
\end{figure}

Next, in Fig,~\ref{fig:mosaic}, we show twelve example late-type
galaxies along with their lopsidedness values in a progression from
symmetric to lopsided.  The three galaxies
shown in the top row each exhibit a regular and symmetric appearance
and have low values of lopsidedness $A_1 \le 0.04$. The next 5
galaxies have moderate $A_1$ values of 0.08$-$0.13, and minor
distortions are visible in each case.  In cases of two-arm spiral
galaxies, one or both arms are disturbed so that their shape and
brightness are not identical at equal distances from the galactic
center.  In the case of SDSS J084845.62+001729.5 (third row, left), a
second arm on the right has no clear counterpart on the left,
and the lopsidedness of this galaxy is 0.11$-$0.12 depending on the
band.  The final four galaxies shown have clear asymmetry in
brightness and shape and have $A_1 > 0.16$ in each band.

\begin{deluxetable}{l|l|lrrr}
\tabletypesize{\small}
\tablecolumns{6}
\tablewidth{0pc}
\tablecaption{Example Galaxies in Figure~\ref{fig:mosaic}}
\tablehead{
	\colhead{Row} &
	\colhead{Column} &
	\colhead{Name}  &
   \colhead{$A_1^g$} & 
	\colhead{$A_1^r$} & 
	\colhead{$A_1^i$} }
\startdata
 & Left & J101620.36+542847.1 & 0.04 &  0.04 & 0.02 \\
1st & Middle & J121935.76+055048.3 &  0.04 & 0.03 & 0.03 \\
 & Right & J083104.39+393721.9 & 0.03 & 0.04 & 0.04 \\
\hline
 & Left & J155547.64+423625.9 & 0.08 & 0.07 & 0.06 \\
2nd & Middle & J103705.40+124614.0 & 0.08 & 0.09 & 0.08 \\
 & Right & J225422.30$-$101025.8 & 0.11 & 0.10 & 0.10 \\
\hline
& Left & J084845.62+001729.5 & 0.12 & 0.12 & 0.11 \\
3rd & Middle & J085747.32$-$001159.9 & 0.13 & 0.13 & 0.13 \\
& Right & J232923.56+141215.4 & 0.18 & 0.16 & 0.16 \\
\hline
& Left & J124944.67$-$025826.4 & 0.20 & 0.20 & 0.22 \\
4th & Middle & J144930.14+610654.2 &  0.27 & 0.27 & 0.25 \\
& Right & J094706.75+541842.8 & 0.26 & 0.27 & 0.26
\enddata
\label{tab:mosaic}
\end{deluxetable}

\section{Final Sample Selection: The Effect of Systematic Errors \label{sec:syserr}}

Lopsidedness is ideally measured from an image of a bright, well
resolved, and face-on galaxy without any overlapping background or
foreground sources.  In such an image, the Poisson noise of the
detected light is negligible, and the size of the point-spread
function is small compared to the characteristic size scale of the
lopsidedness. Neither dust extinction nor inclination would
significantly alter the light distribution.  In real data, these
conditions are not all met, and the resulting effects can
systematically change the measurements of the Fourier modes.

Below we address in turn how various sources of systematic error
affect the strengths of Fourier modes.  We then use this information to refine our sample selection and define a final sample of galaxies whose images are only negligibly affected by these systematic errors. These will form the basis of our scientific analysis.

\subsection{Ellipticity and Inclination \label{sec:inclination}}

We employ a circular polar grid to bin light from each galaxy, even if
the galaxy appears elliptical and/or inclined on the image.  The
measured lopsidedness of elliptically projected galaxies can be
underestimated by using the circular grid if the deprojected bright side of the
galaxy coincides with the {\em minor} axis of the galaxy.  This occurs
because the increment and decrement of light along the minor axis due
to lopsidedness is compared to the average light along a circular ring
that also intersects the brighter, inner region at the major axis.
Similarly, the lopsidedness will be overestimated if the bright and dim
sides of the galaxy are aligned with the {\em major} axis.  The
resulting systematic error is diminished if the phase angle of the
first Fourier mode wraps at least a quarter of the way around the
galaxy.  In this case, the first mode strength would be overestimated
at some radii and underestimated at other radii, and the radially
averaged $A_1$ would have a smaller systematic error.

\begin{figure}
\epsscale{1.1}
%\plottwo{/home/tar/agn/asymm/lophists/meddist_baratio_a1i.ps}{/home/tar/agn/asymm/lophists/meddist_baratio_a2i.ps}
\plottwo{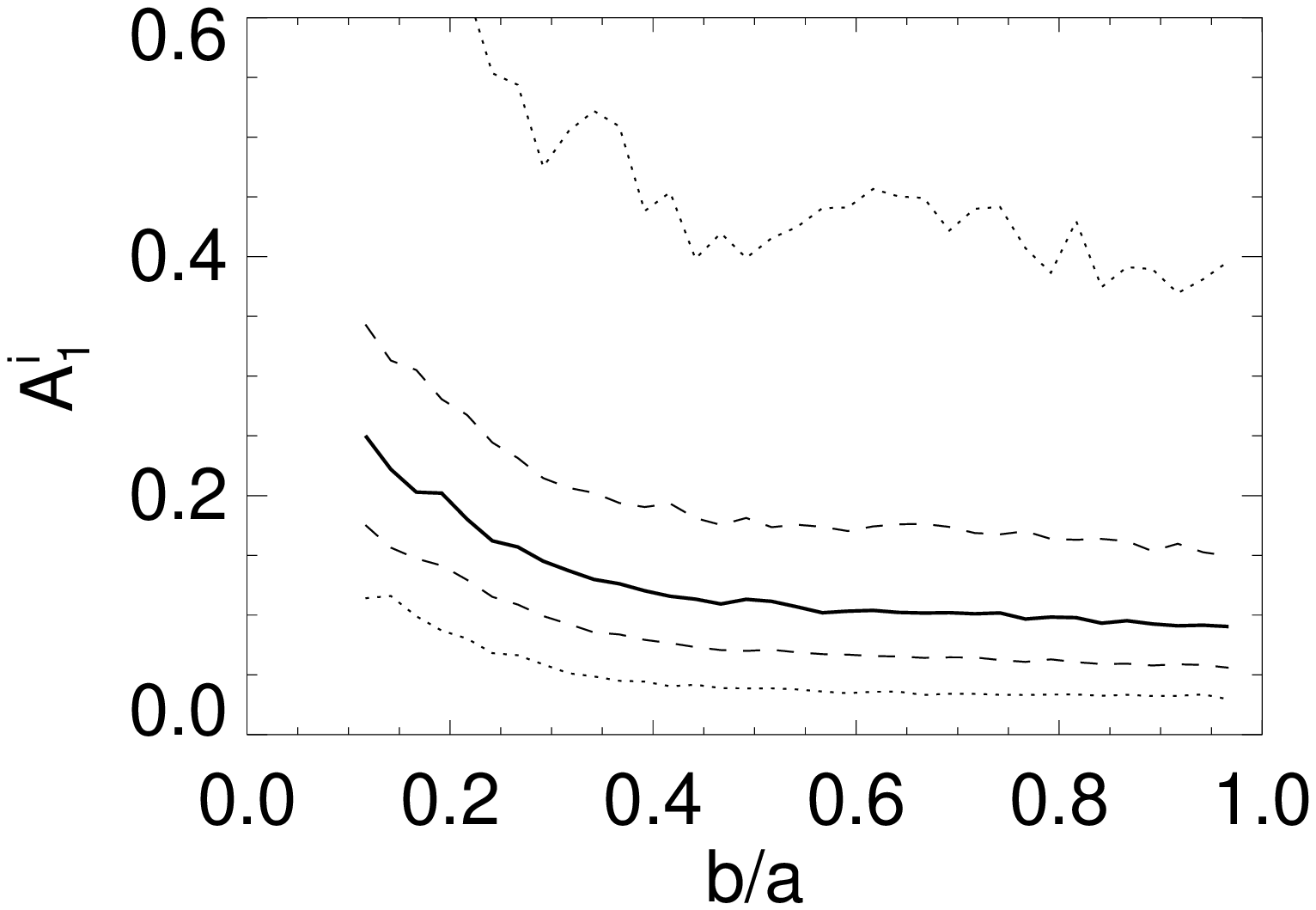}{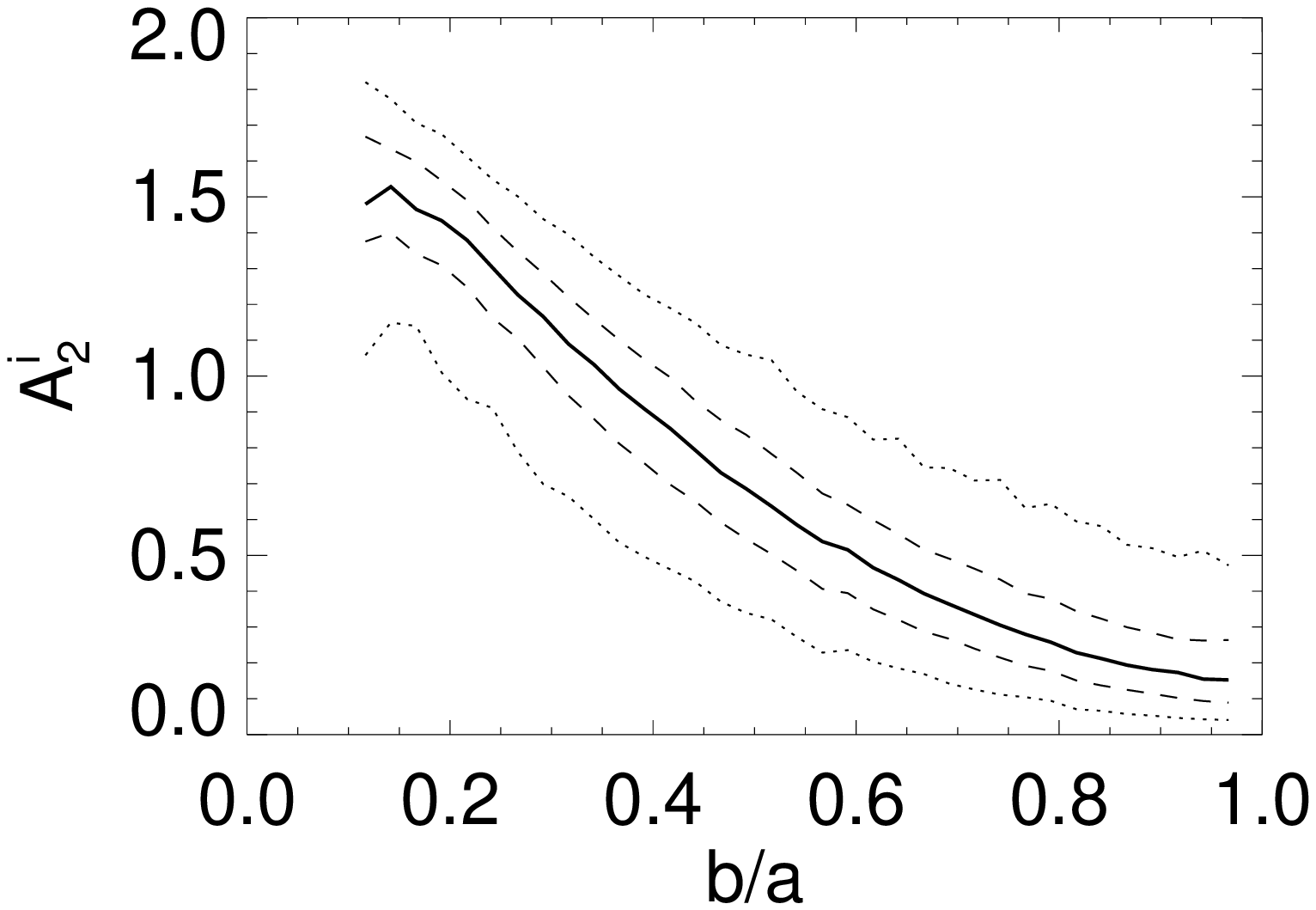}
%\plottwo{/home/tar/agn/asymm/lophists/meddist_baratio_a3i.ps}{/home/tar/agn/asymm/lophists/meddist_baratio_a4i.ps}
%\plottwo{/home/tar/agn/asymm/lophists/meddist_baratio_a5i.ps}{/home/tar/agn/asymm/lophists/meddist_baratiogri_hist.ps}
\epsscale{0.5}
%\plotone{/home/tar/agn/asymm/lophists/banddist_baratio.ps}

\plotone{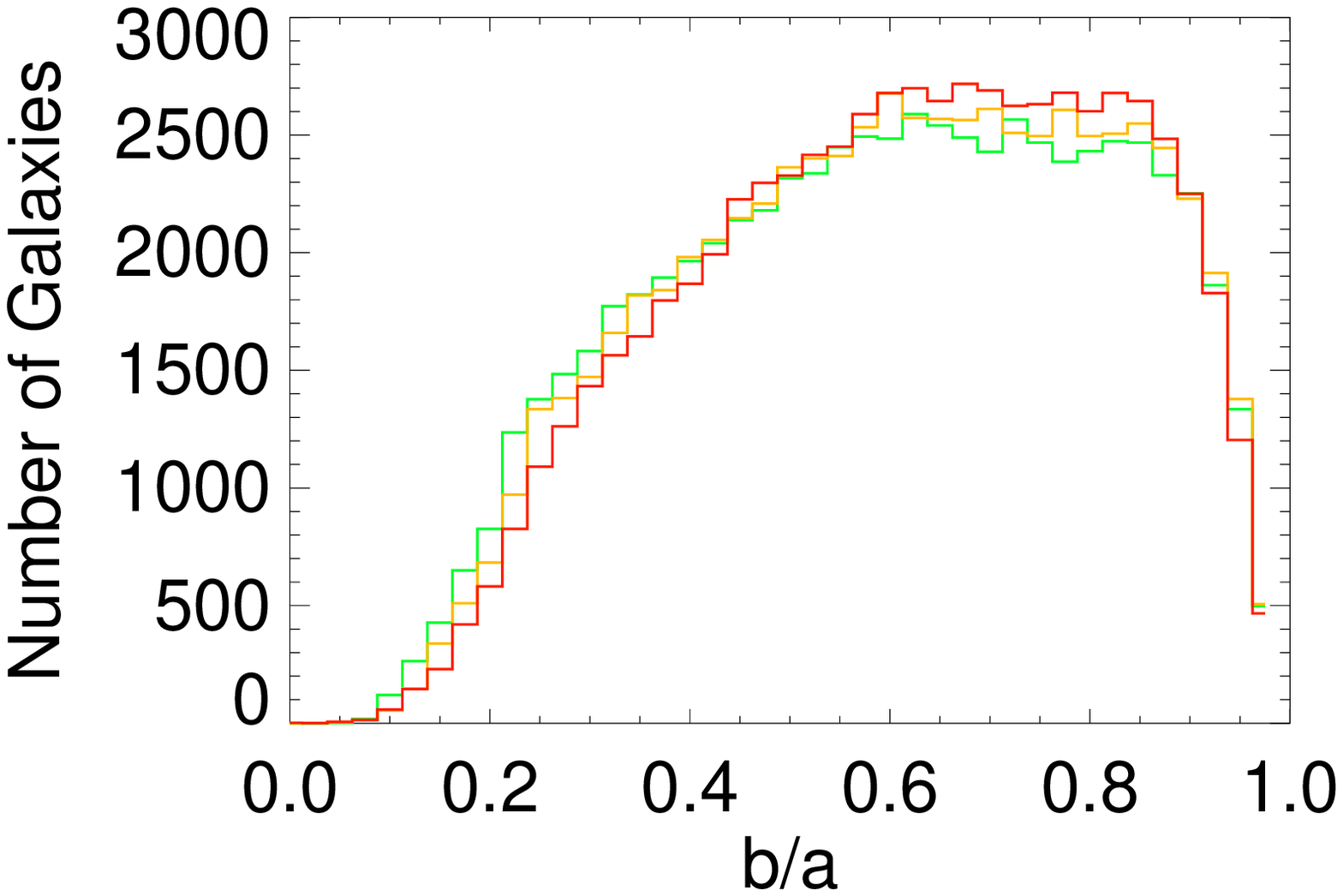}
\caption{ {\em (Top two panels)} Distributions of the Fourier modes $A_1$ and $A_2$ as functions of $b/a$ in the $i$-band.  The 5th, 25th, 50th, 75th, and 95th percentiles are shown.  Lopsidedness is independent of inclination for $b/a > 0.4$.  On the other hand, $A_2$, an indicator of ellipticity, inclination, and two-arm spiral patterns, decreases significantly with $b/a$ along the full range of $b/a$.  {\em (Lower right)} The similar distributions of $b/a$ as measured in the $g$- {\em (green)}, $r$- {\em (orange)}, and $i$-bands {\em (red)}.  A three-color cut at $b/a = 0.4$ eliminates only a small portion of the sample.}
\label{fig:lopba}
\end{figure}

Our sample has galaxies imaged in random orientations, and the over-
and under-estimates of lopsidedness average to cancel each other
except for highly inclined galaxies.  We show this result in
Fig.~\ref{fig:lopba}, where we have plotted the distribution of
$A_1^i$ (and also of $A_2^i$) as a function of $b/a$ as measured in
the $i$-band (similar results are seen in the $g$ and $r$-bands).

The relationship between $b/a$ and $A_2^i$ is clear.  Face-on galaxies
have a small second mode, typically 0.1-0.3, while highly inclined
galaxies have a much higher second mode that often exceeds unity for
$b/a < 0.4$.  Though not shown, higher-order even modes show a similar
trend with $b/a$ as $A_2^i$ but with weaker magnitude.

Lopsidedness exhibits an increase at low $b/a$ but is mainly
independent of $b/a$ for $b/a > 0.4$.  The rise of observed
lopsidedness for increasingly inclined galaxies arises for a few
reasons.  First, dust lanes appear more optically thick when a galaxy
is viewed nearly edge-on. A lopsided, edge-on distribution of dust may
obscure the light in a symmetric galaxy to make the galaxy light
appear lopsided.  Second, the systematic error in using a circular
grid for an elliptically projected galaxy becomes more severe at low
$b/a$ ratio.  Nonetheless, $A_1^i$ shows little dependence on $b/a$
for $b/a > 0.4$, so our use of a circular polar grid and a
finite-order transform on slightly elongated galaxies produces no
significant systematic error when in reference to a large population
of galaxies. The distributions of $b/a$ as measured in the $g$-, $r$-,
and $i$-bands are given in the lower panel of Fig.~\ref{fig:lopba}.  A
cut at $b/a = 0.4$ in all three bands eliminates the 22\% of the
sample, and 52194 galaxies are retained. This cut on inclination is very similar to that adopted by \citet{bc+05}.

\clearpage

\subsection{Spatial Resolution \label{sec:blur}}

Lopsided galaxies systematically appear more symmetric when observed
in conditions of poor seeing.  Light is smeared from the brighter
regions of the galaxy into surrounding, dimmer regions, reducing the
contrast of the brighter and dimmer sides of the galaxy.  The effect
is more pronounced when the point-spread function is wide compared to
the physical size of the galaxy.  It is useful to define a seeing
resolution
\begin{equation}
S_b \equiv \frac{2R_{50,b}}{{\rm PSF\,\; FWHM}_b}
\end{equation}
as the relative size of the galaxy compared to the FWHM of the PSF in
band $b$.  

\begin{figure}[ht]
\epsscale{1.0}
%\plotone{/home/tar/agn/asymm/degrade.hist_res_a1bins.ps}
\plotone{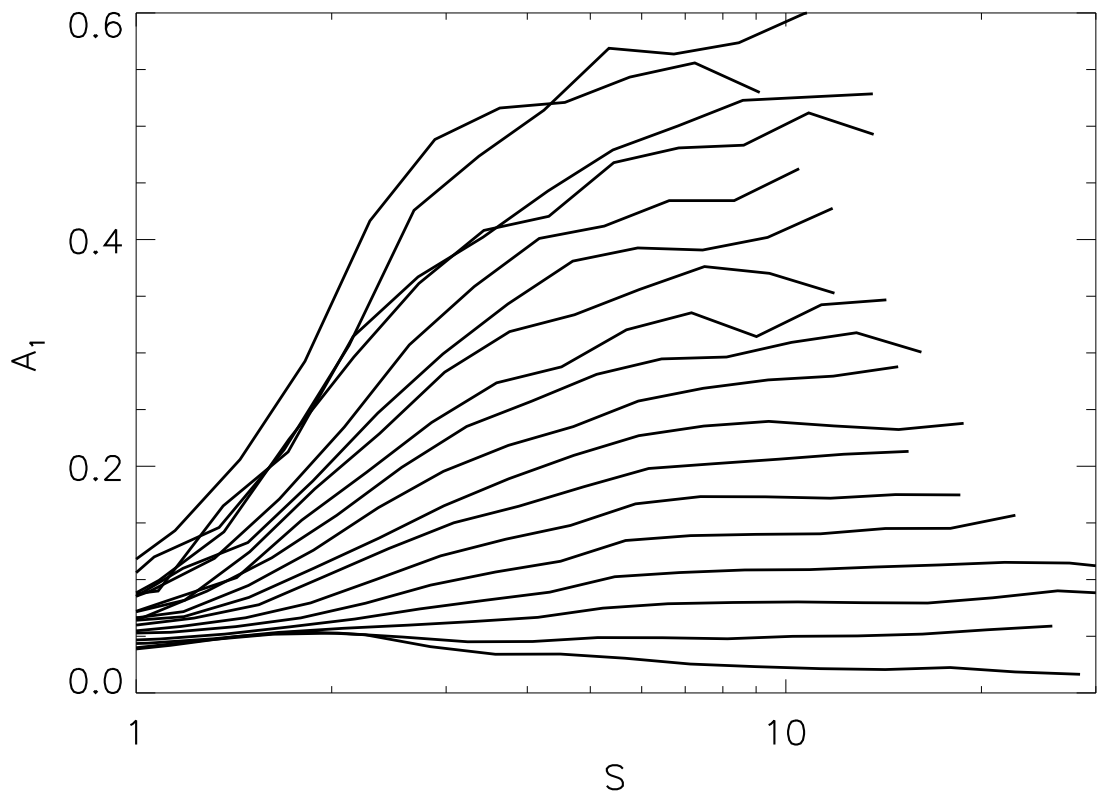}
\caption{The $i$-band lopsidedness of galaxies in images simulating various degrees of seeing resolution.  A sample of isolated, well-resolved galaxies was divided into bins by $A_1$ (measured before blurring) and the images were then blurred.  For each bin, the relation between median $A_1$ after blurring and the ratio of the half-light and blurred PSF diameters ($S$) is shown above.  Lopsided and symmetric galaxies can be distinguished for $S > 3$ but the range of $A_1$ is diminished for $S \lesssim 4$.  } 
\label{fig:psfblur}
\end{figure}

To demonstrate how poor seeing reduces lopsidedness, we
have selected from our galaxy sample a subsample of 13500 isolated,
well resolved galaxies with $b/a > 0.5$ and $S_i > 5.0$ and convolved
the $i$-band images with a circular Gaussian PSF of 20 varying widths.
We calculated $A_1^i$ of each galaxy at each level of blurring.  The
subsample was binned by $A_1^i$ of the original (not blurred) images.
Fig.~\ref{fig:psfblur} shows how the median $A_1^i$ of the blurred
images changes with seeing resolution in each bin of the original
$A_1^i$.  Post-blurring $A_1^i$ exhibits little change for highly
resolved $S > 5$ galaxies.  At moderate resolutions $2 < S < 5$,
lopsidedness decreases with the seeing resolution.  Barely resolved
galaxies with $S < 2$ approach $A_1^i \sim 0.05$ regardless of the
lopsidedness at high resolution.

\clearpage

\begin{figure}[ht]
\epsscale{1.0}
%\plottwo{/home/tar/agn/asymm/lophists/meddist_a1i_resi.ps}{/home/tar/agn/asymm/lophists/banddist_res.ps}
\plottwo{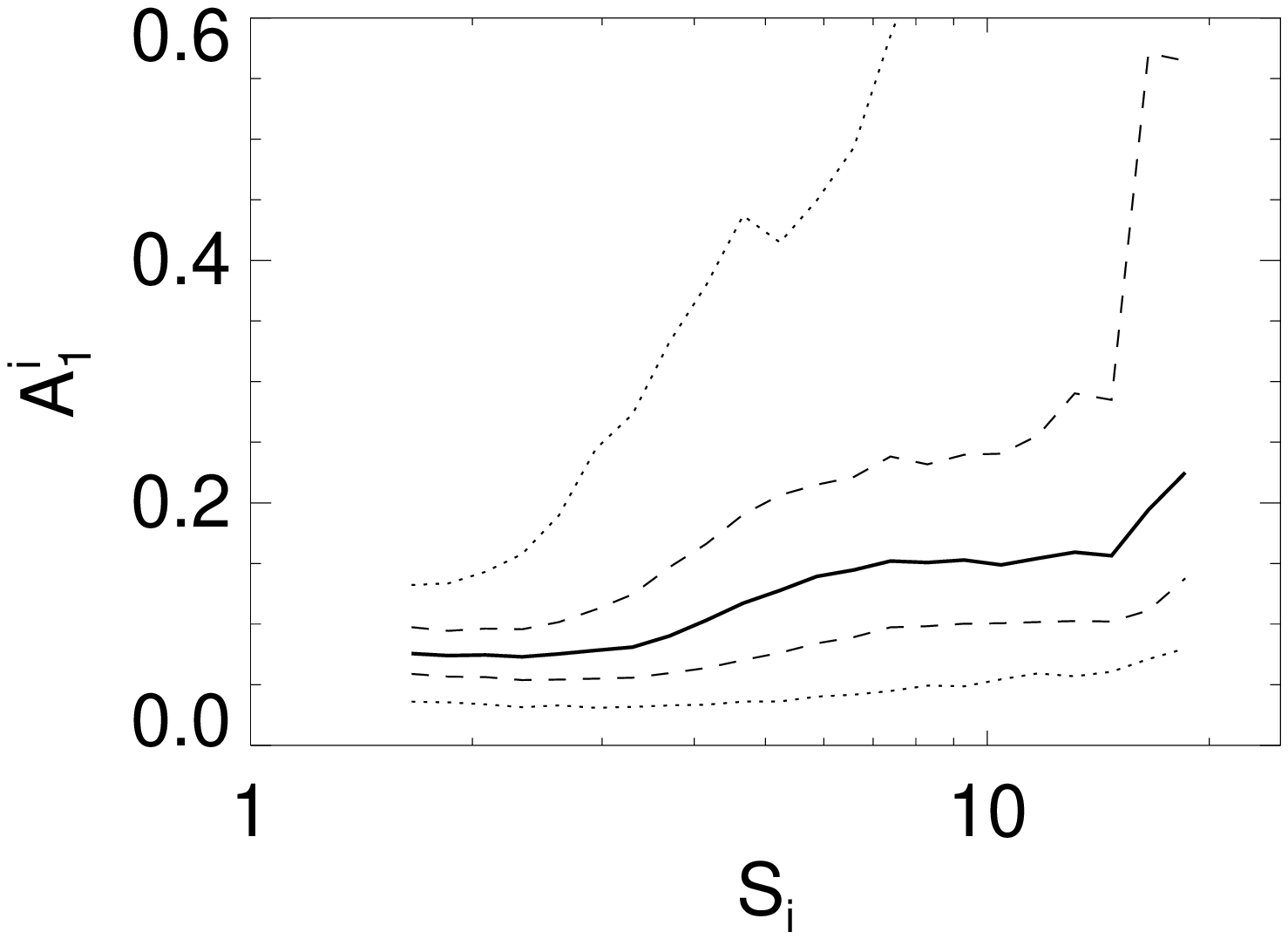}{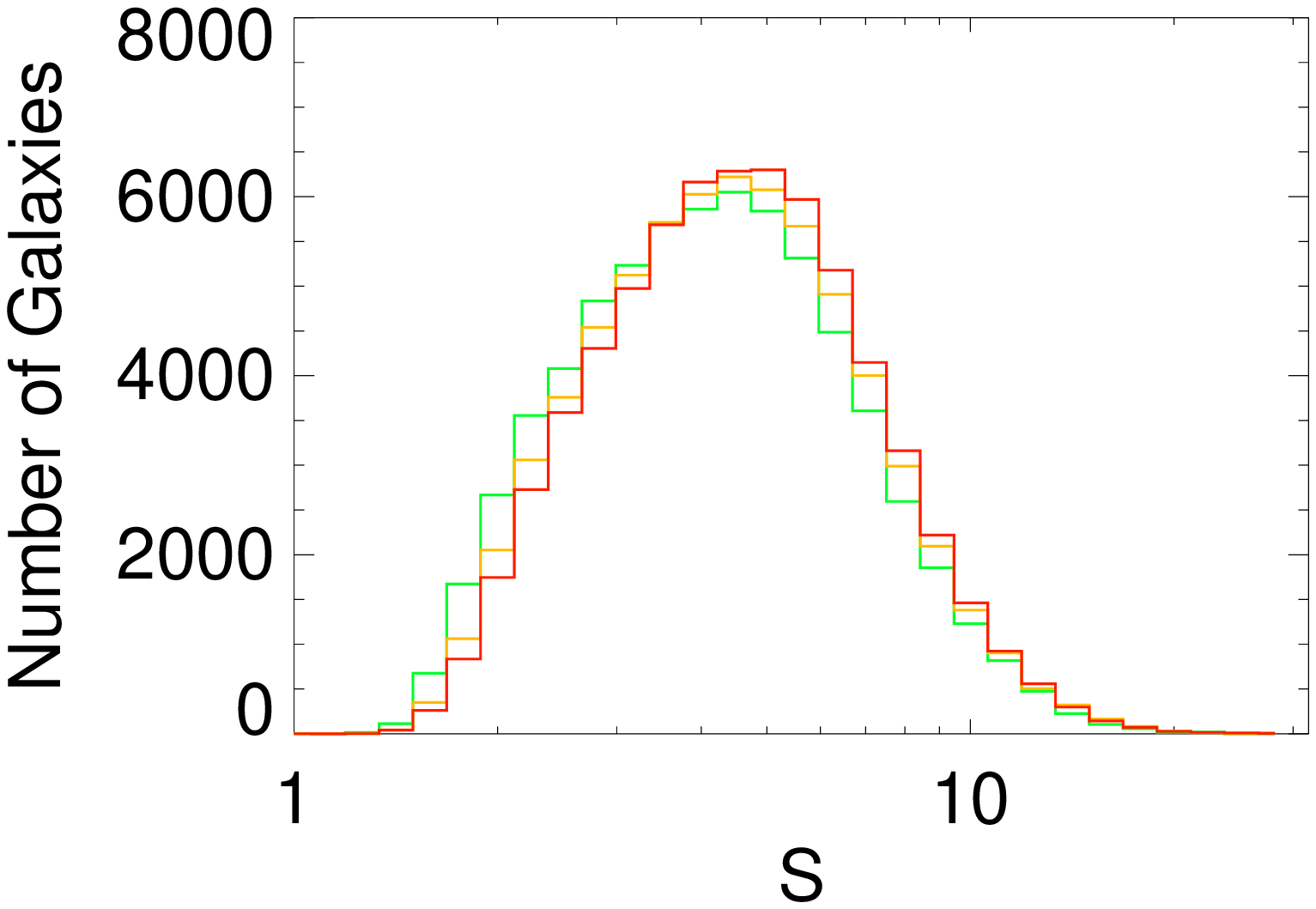}
\caption{{\em Left:} Distribution of $i$-band lopsidedness for galaxies observed at different seeing resolution. The 5th, 25th, 50th, 75th, and 95th percentiles are shown.  {\em Right:} Distribution of seeing resolution as measured in the $g$-band {\em (green)}, $r$-band {\em (orange)}, and $i$-band {\em (red)}.  Blurring from poor seeing systematically reduces the measured lopsidedness.  A cut in seeing resolution at $S = 4$ removes 37\% of the sample while allowing a distinction between symmetric and asymmetric galaxies.}
\label{fig:psfdist}
\end{figure}

The same effect is seen in our primary sample of galaxies without
convolving the images with a point-spread function.
Fig.~\ref{fig:psfdist} shows the distribution of $A_1^i$ at different
seeing resolutions along with the distributions of $S_g$, $S_r$, and
$S_i$ for the whole sample (before the inclination cut is made).  At
low seeing resolution $S < 3$, the sample distribution of $A_1^i$
shifts to lower lopsidedness (0.07 median), and few galaxies have
$A_1^i > 0.2$.  At high resolution $S > 7$, the distribution has a
median that levels off around 0.12, and high lopsidedness $A_1^i >
0.20$ is found in 1 of every 4 galaxies.  The 5th percentile increases
with resolution to $A_1^i = 0.05$ at $S = 7$.

Our simulations of poor seeing suggest that a distinction between
symmetric and lopsided galaxies can be identified by the $A_1$ measure
even if $A_1$ is diminished by moderately poor seeing.  The relative
lopsidedness of well-resolved galaxies is preserved as the seeing is
worsened down to $S \sim 3$, i.e., the most symmetric galaxies with $3
< S < 7$ are likely to also be the most symmetric galaxies if they
were observed instead at $S > 7$, and likewise for the more lopsided
galaxies.  However, the extremely lopsided galaxies may not be able to
identified without higher resolution, perhaps $S > 10$.  Trends at
extreme lopsidedness $A_1 > 0.3$ may not be reliable because only well
resolved galaxies can lie in this lopsidedness range.  At $S = 4$,
seeing reduces the lopsidedness of 75\% of galaxies by less than a
factor of 1.5. We cut our sample at this seeing value and keep only
galaxies better resolved than $S = 4$ in the $g$, $r$, and $i$ bands.
The cut alone reduces the sample size by $\sim 37\%$ to 42558
galaxies.  The more lopsided galaxies $A_1 > 0.2$ are underrepresented
due to this systematic effect.

%Because
%information about lopsidedness is not lost for moderately resolved
%galaxies, we can attempt a correction to $A_1$.  In
%Fig.~\ref{fig:blurcorr}, we show, as a function of $S$, the ratio of
%the observed lopsidedness after blurring to the average lopsidedness of
%simulations of the same galaxy with $S > 8$.  While there is some
%scatter in the lopsidedness ratio, we can use the median ratio as a
%correction to $A_1$.  [**I'm presenting the correction for now.  I
%haven't used it anywhere else yet. Do I even need to show it?]

\clearpage

%\begin{figure}[ht]
%\epsscale{1.0}
%\plotone{/home/tar/agn/asymm/degrade.correction.ps}
%\caption{The correction factor for moderately resolved galaxies.  The vertical axis shows the ratio of the lopsidedness of a blurred galaxy to the lopsidedness the same galaxy has when less blurring is applied ($S > 8$).}
%\label{fig:psfdist}
%\end{figure}
%\clearpage

\subsection{Flat-fielding error} 

Poor flat-fielding in a galaxy image can increase the apparent
lopsidedness of the galaxy.  Poor correction of the CCD's spatial
gradients in sensitivity can introduce a brighter sky level on one
side of the galaxy and a dimmer sky level on the other side.  Because
a constant sky value is subtracted from the image in the lopsidedness
computation, the gradient in sky brightness will increase the strength
of the odd-order Fourier modes and overestimate the galaxy's true
lopsidedness.  We have calculated flat-field errors in our images by
calculating the sky level in regions in each of the four corners of
the image and adopting the standard deviation of those sky levels as
our flat-fielding error.  We find that flat-fielding errors are
typically negligibly small, $\sim 1$\% of the sky value, and only
become significant at large radii from a galactic center.  In our
calculation of $A_m$, the contribution of the modes at large radii is
weighted less, and so flat-fielding errors do not significantly affect
the lopsidedness values that we have computed.  We therefore do not
impose a cut based on flat-fielding error for our SDSS-detected
galaxies.

\subsection{Random Error \label{sec:randerr}}

Noisy data from low-surface-brightness galaxies can bias measurements
of lopsidedness in a systematic way. The uncertainty in the correct
light-weighted center will be higher in a dim galaxy than in a bright
one, and a shifted center leads to an overestimate of the
lopsidedness.

\begin{figure}[ht]
\epsscale{1.0}
%\plottwo{/home/tar/agn/asymm/lophists/meddist_sni_a1ri.ps}{/home/tar/agn/asymm/lophists/banddist_sn.ps}
\plottwo{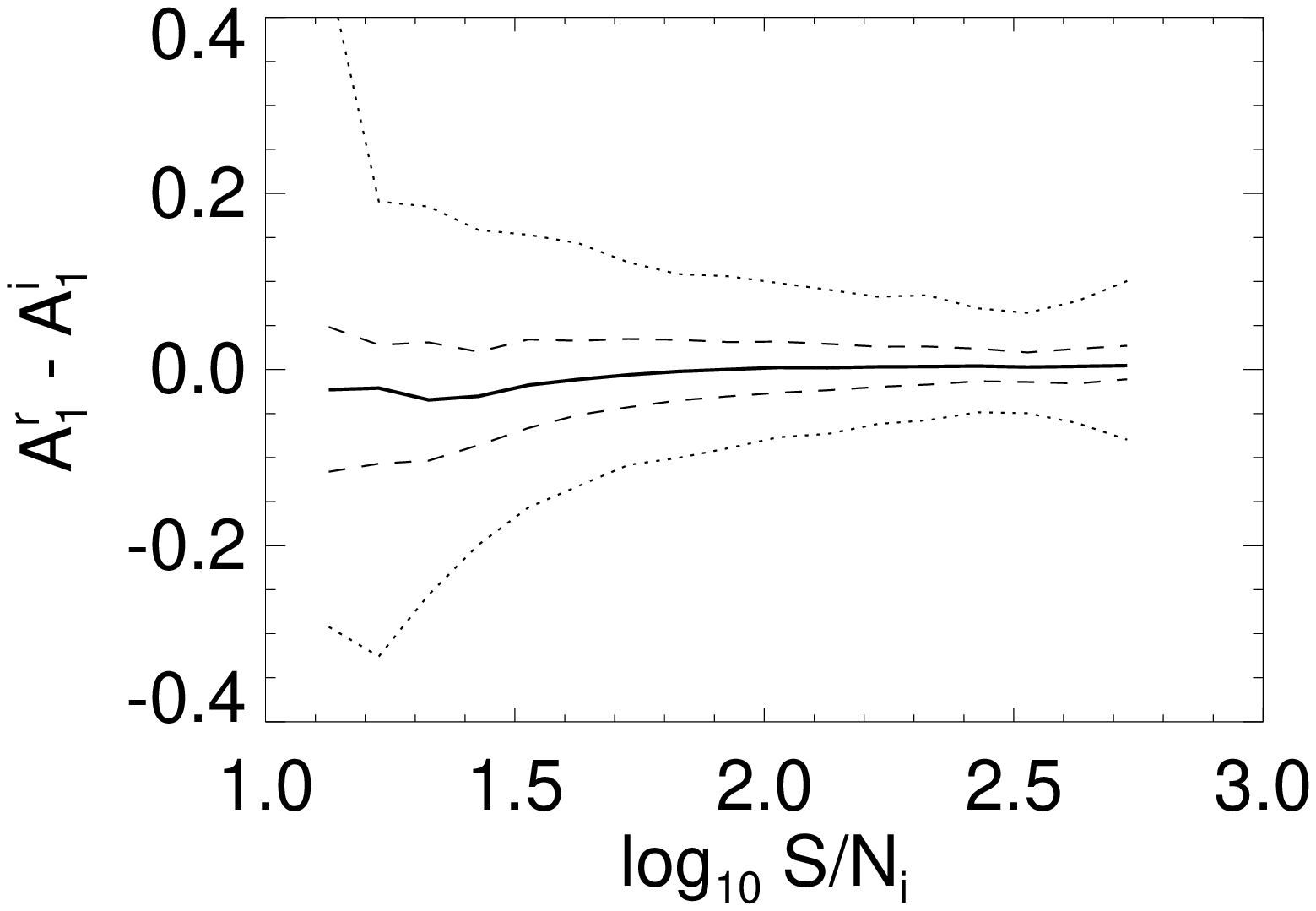}{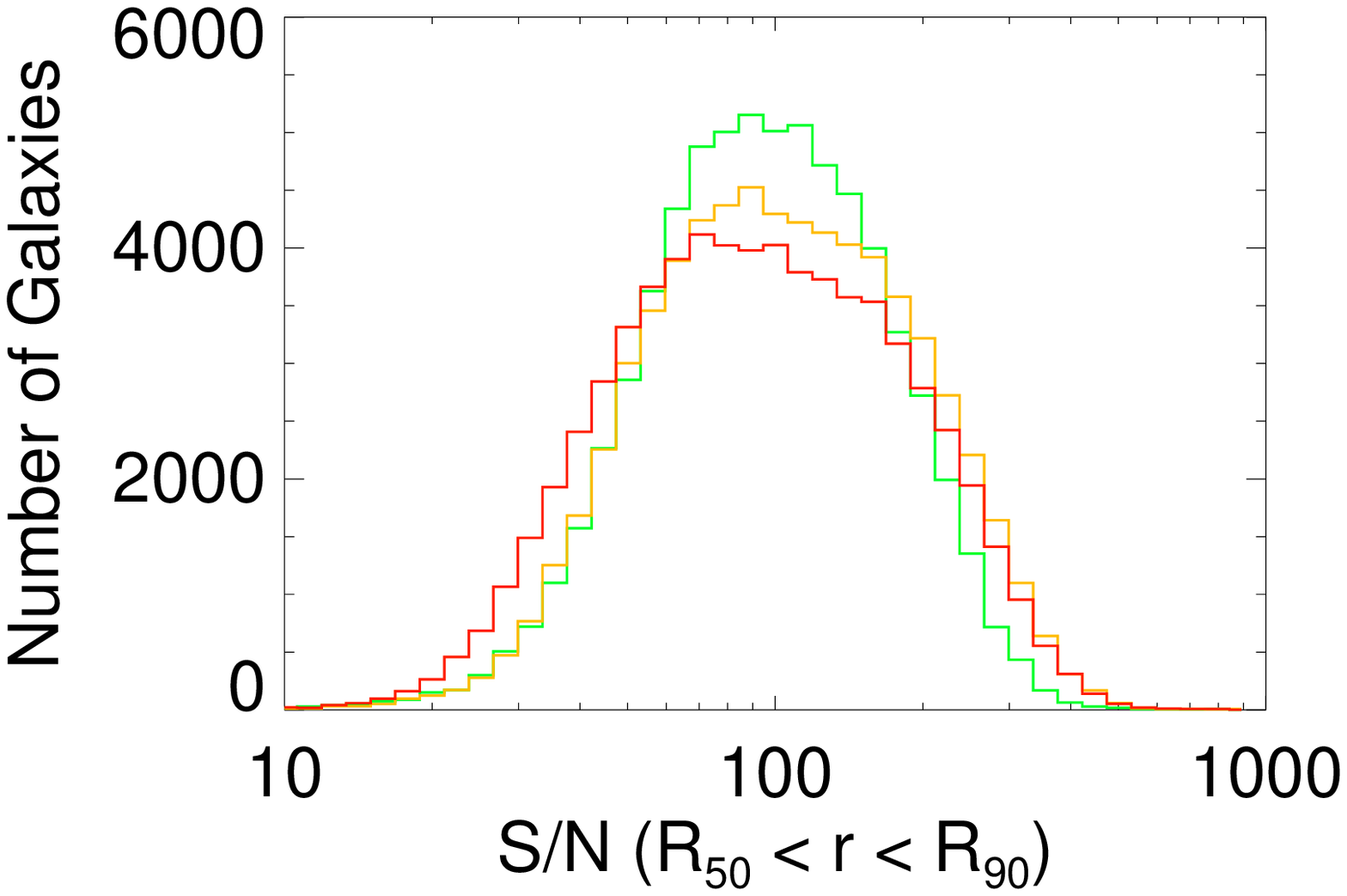}
\epsscale{1.1}
\caption{{\it (Left)} Distributions of the difference in lopsidedness in the $r$ and $i$ bands as a function of $i$-band S/N.  The 5th, 25th, 50th, 75th, and 95th percentiles are shown.  The difference is typically small ($|A_1^r-A_1^i| < 0.04$) for most of the sample with $S/N_i > 30$. Since lopsidedness is systematically similar in these two bands, random errors in $A_1^r$ and $A_1^i$ should be of the similar size as the difference $|A_1^r-A_1^i|$.  {\it (Right)} The distribution of $S/N$ in the $g$- (green), $r$- (orange), and $i$-band (red).  A cut at $S/N < 30$ in all three bands eliminates a small fraction of the sample. } 
\label{fig:lopri}
\end{figure}

We have undertaken two tests to determine the effect of noise on our
measurements. First, we have compared the values of $A_1$ measured in
the $r$- and $i$-bands. Lopsidedness values have been
determined before to be mainly independent of wavelength in this range
\citep{rr98}, and we will confirm this result for our sample as a
whole in \S\ref{sec:lightmass}.  In the left panel of Fig.~\ref{fig:lopri},
we show the discrepancies between $A_1$ in the two bands as a rough
indicator of random error on the $A_1$ measurement in general.  Here
the $S/N$ ratio is computed for light within the $R_{50}$-to-$R_{90}$
annulus in which $A_1$ is computed.  The interquartile (25th-75th
percentile) range is small at high $S/N$ ($\lesssim 0.04$ at $S/N =
300$) and gradually increases at low $S/N$ as the distribution spreads
out.  At $S/N = 30$, the interquartile range has doubled to $0.09$, and the
median is skewed only slightly toward negative values of $A_1^r-A_1^i
\sim -0.01$.  

We have applied another cut to our sample to remove dim, noisy
galaxies with $S/N < 30$ in any of the three $gri$ bands.  The right
panel of Fig.~\ref{fig:lopri} shows the distribution of $S/N$ in three
bands.  The cut removes a 5\% of the sample.  The sample retains 25155
galaxies after this cut and also the inclination and seeing cuts have
been made.

\clearpage

To further evaluate the effect of noise in calculating lopsidedness,
we looked to the full SDSS DR4 dataset to retrieve a $z < 0.06$ sample
of galaxies that have been observed 3 or more times.  We narrowed this
sample down to a smaller sample where inclination, seeing resolution,
and $S/N$ met the same cuts as our main sample ($b/a > 0.4$, $S > 4$,
and $S/N (R_{50} < r < R_{90}) > 30$). We also required that the
fractional RMS variations between different observations in seeing
resolution and $S/N$ were at most 10\%.  This latter requirement
ensures that the repeated observations had similar observing conditions.
The resulting sample contained 328 galaxies with repeated
observations. We calculated the RMS differences in $A_1$ in the
repeated observations of these galaxies and adopted this measure as
the error $\delta A_1$ in lopsidedness for these galaxies.

\begin{figure}[ht]
\epsscale{1.0}
%\plotone{/home/tar/agn/asymm/lophists/reobs_lopc_dlop.ps}
\plotone{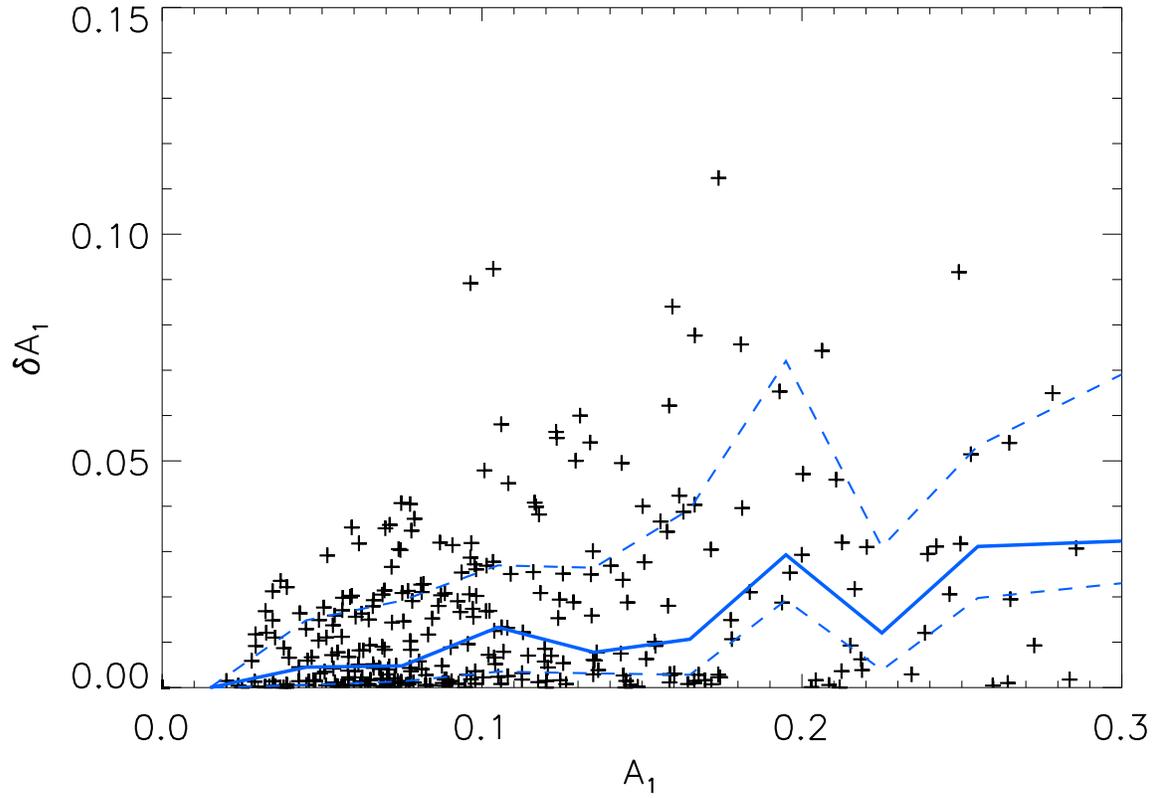}
\epsscale{1.1}
%\plottwo{/home/tar/agn/asymm/lophists/reobs_a2_da2.ps}{/home/tar/agn/asymm/lophists/reobs_a3_da3.ps}
%\plottwo{/home/tar/agn/asymm/lophists/reobs_a4_da4.ps}{/home/tar/agn/asymm/lophists/reobs_a5_da5.ps}
\caption{Errors in lopsidedness for galaxies with repeated observations. The 25th, 50th, and 75th percentiles are over-plotted.   Typical lopsidedness  errors are $\sim 0.02$ for the more symmetric galaxies and $\sim 0.1A_1$ for the more lopsided galaxies.  }
\label{fig:randerr}

\end{figure}

Fig.~\ref{fig:randerr} shows the distribution of $\delta A_1$
vs. $A_1$ for the repeatedly observed galaxies.  About a quarter of
the galaxies show small errors in lopsidedness, $\delta A_1 < 0.01$,
and median errors are typically $\sim 0.1A_1$.  The largest errors are
$\sim 0.5A_1$ but affect only 5\% of the sample. The third quartile
rises as $\sim 0.2A_1$ but can be as high as 0.02 at lopsidedness as
low as 0.05.  Typical errors can be expected to be the larger of 0.02
and $0.10A_1$.  The majority of the sample has $A_1 < 0.20$, and so
0.02 can be taken as the typical error for most of the sample, with
larger errors for the most lopsided galaxies. This result is
consistent with our estimate above based on comparing the $r$-band and
$i$-band values for a much larger sample.

\clearpage

\subsection{A Sample Suitable for Calculating Lopsidedness \label{sec:cuts}}

We have discussed the systematic errors that skew our calculation of
lopsidedness.  Here we summarize the proposed cuts to weed out cases
where the systematic effects give unphysical values of $A_1$.  After
applying these cuts, we are left with a significant sample that is
suitable for studying correlations between lopsidedness and other
global properties of galaxies.

\begin{itemize}

%\item {\em Overlapping galaxies.}  We eliminate all galaxies in our sample that overlap another galaxy brighter than 6 apparent magnitudes fainter than our sample galaxy.  Galaxies with only faint overlapping galaxies and isolated galaxies remain in the sample.

\item {\em Inclined galaxies.}  We eliminate all galaxies 
with $b/a < 0.4$ in any of the three bands to remove inaccurate $A_1$
values computed using a circular aperture on an elliptically projected
galaxy. This cut also removes cases of obscuration from optically thick
dust lanes on edge-on galaxies.

\item {\em Poorly resolved galaxies.} We eliminate all galaxies 
with seeing resolution $S < 4$ in any of the three bands in order to
remove galaxies whose lopsidedness is diminished due to poor seeing.

\item {\em Dim galaxies.}  We eliminate all galaxies with $S/N < 30$ 
in the $R_{50}$-to-$R_{90}$ annulus (the region where $A_1$ is
calculated) in any of the three bands to remove cases where
lopsidedness is augmented by Poisson noise and poor centering.

\end{itemize}

Table~\ref{tab:cuts} shows the census of galaxies retained after each
cut is applied separately.  The resolution cut alone removes the
largest portion of the sample, and the noise cut removes the least
amount.  After all three cuts are applied to the initial sample of
67107 galaxies, 25155 (37\%) are retained. Unless otherwise specified,
we have employed these cuts in our sample for all analysis presented
below.  These cuts depend on parameters linked to the observation of
the galaxies but may also depend on physical properties.  We next look
at the structural properties of the sample and compare them to the
larger DR4 sample from which it was drawn.

\begin{deluxetable}{lrr}
\tabletypesize{\small}

\tablecolumns{3}
\tablewidth{0pc}
\tablecaption{Sample Cuts}
\tablehead{
	\colhead{} &
	\colhead{Percentage} &
	\colhead{Galaxies} \\
   \colhead{Cut} & 
	\colhead{Retained} & 
	\colhead{Retained} }
\startdata
Initial Sample & 100\%  & 67107 \\
$b/a > 0.4$ & 78\% & 52194 \\
$S > 4$ & 63\% & 42558 \\
$S/N (R_{50} < r < R_{90}) > 30$ & 95\% & 63434 \\
Final Sample after all 3 cuts & 37\% & 25155 \\
\enddata
\label{tab:cuts}
\end{deluxetable}

The main structural parameters we will utilize in the analysis below
are the stellar mass $M_*$, the effective stellar surface mass density
$\mu_*$ (the mean stellar density interior to the z-band half light
radius), and the concentration $C_i$ (defined as the ratio of
$R_{90}/R_{50}$ in the i-band). See \citet{khw+03a} for a detailed
description.

\begin{figure}[ht]
\epsscale{1.1}
%\plottwo{/home/tar/agn/asymm/lophists/banddist_m_cuts.ps}{/home/tar/agn/asymm/lophists/banddist_mu_cuts.ps}
\plottwo{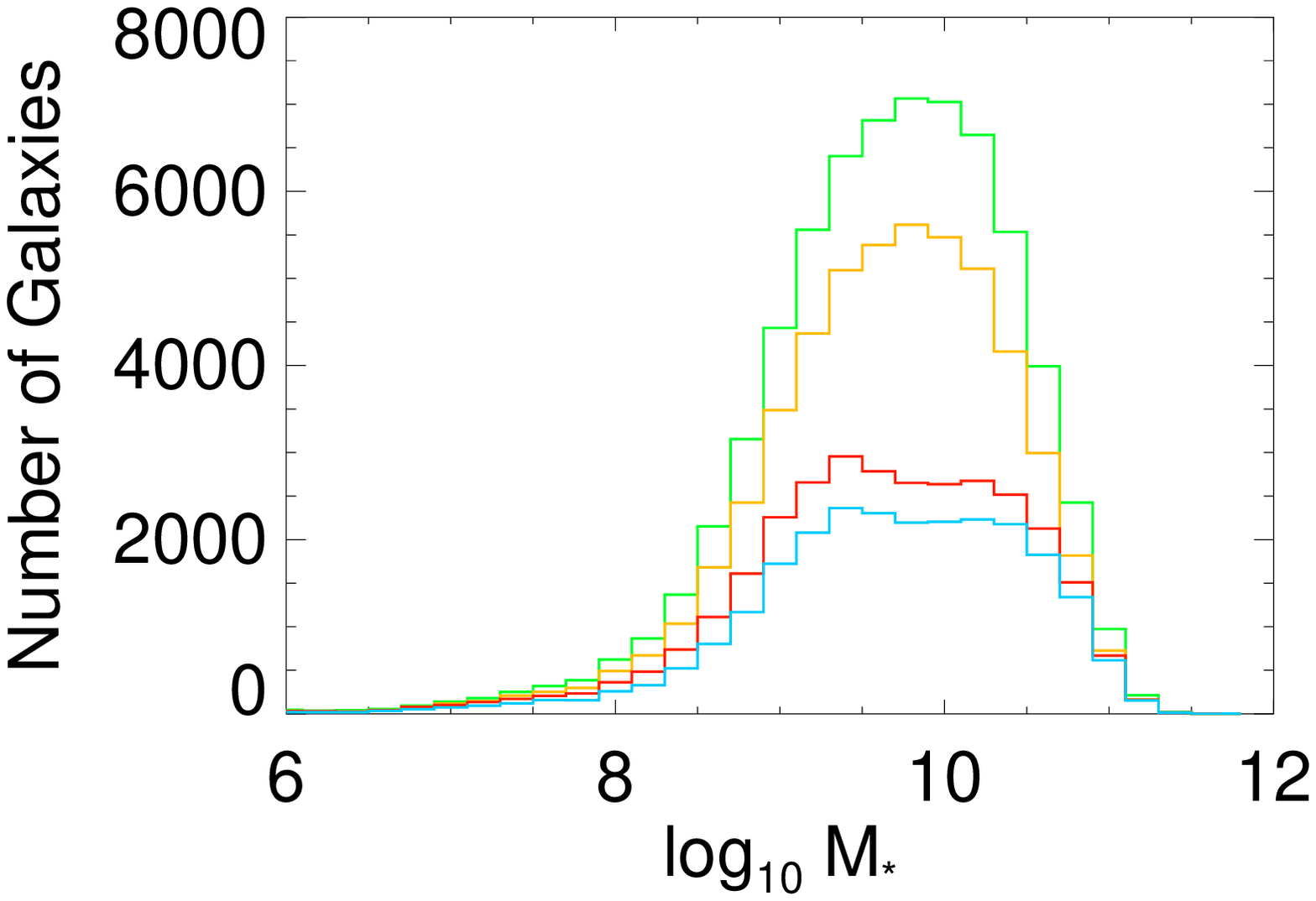}{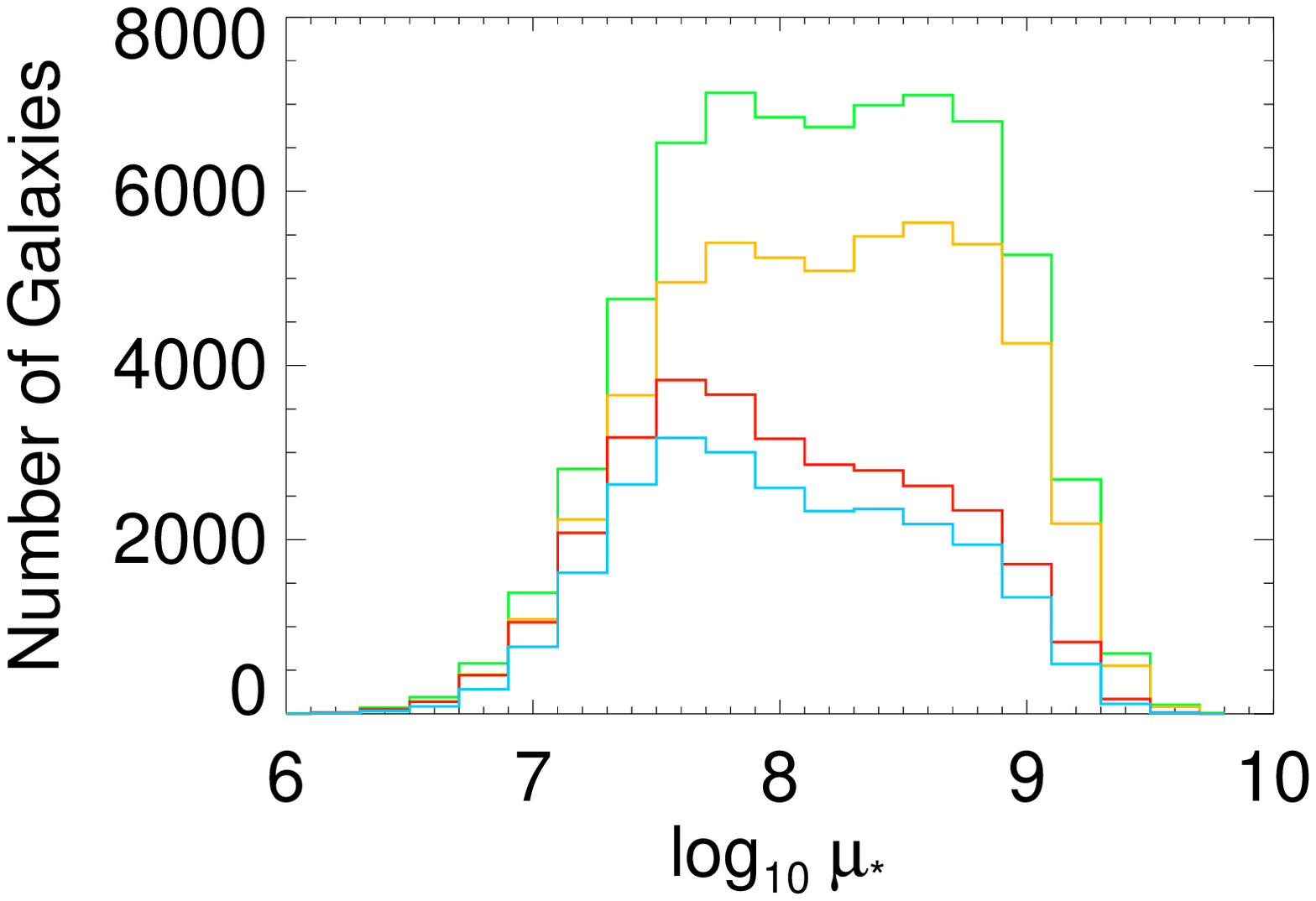}
\epsscale{0.5}
%\plotone{/home/tar/agn/asymm/lophists/banddist_conci_cuts.ps}
\plotone{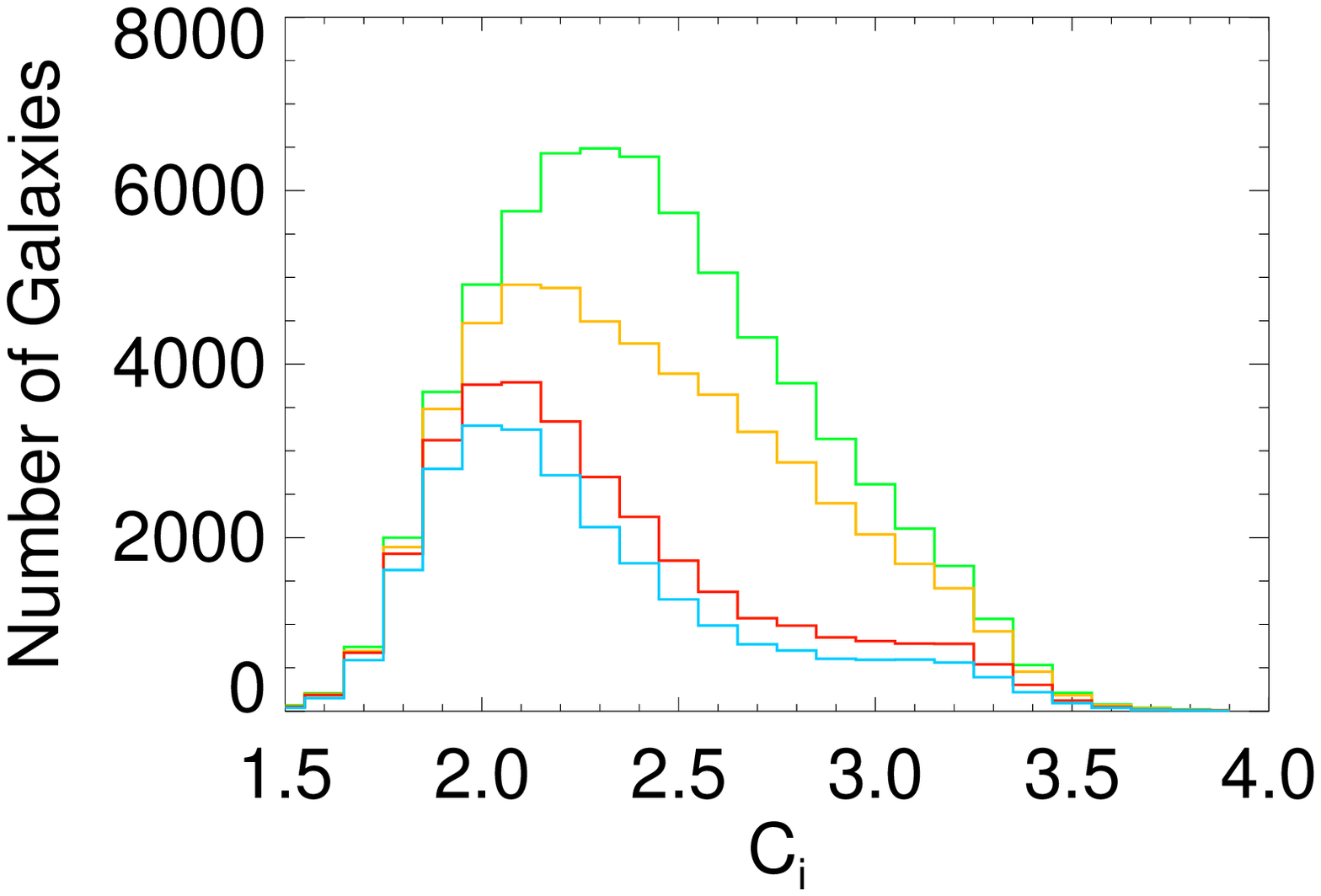}
\caption{Distributions of the structural parameters $C_i$, $M_*$, and $\mu_*$ as the observational cuts were applied.  From top to bottom, in each panel: full DR4 $z<0.06$ sample before cuts were applied; $b/a > 0.4$ cut is applied; $S > 4$ cut is applied; $S/N > 30$ cut is applied.  The resulting sample has similar numbers of low- and high-mass galaxies, low- and high-stellar-density galaxies, but more a larger proportion of late-type galaxies than early-type galaxies.}
\label{fig:lopsamplecut}
\end{figure}

In Fig.~\ref{fig:lopsamplecut}, we show the distributions of these
structural parameters $M_*$, $\mu_*$, and $C_i$ after we apply each
cut in succession to the sample.  The upper curve in each panel shows
the distribution from the initial $z < 0.06$ DR4 sample.  Then the
inclination cut is applied (2nd distribution from the top), followed
by the resolution cut (3rd distribution from the top) and the $S/N$
cut (lower distribution).  The inclination cut reduces the galaxy
counts without any strong correlation with structural properties.  The
resolution cut rejects more massive, high-mass-density, and
concentrated galaxies, leaving the sample with a preference of
late-type galaxies.  Finally, the $S/N$ cut causes little change in
the relative proportions of massive, high-mass-density, or
concentrated galaxies.

The final sample contains 25155 galaxies spanning 3 orders of
magnitude in stellar mass ($10^8-10^{11} \MSun$), 3 orders of
magnitude of stellar mass density ($10^{6.5}-10^{9.5} \MSunkpc$), and
a wide range of $i$-band concentration ($1.5-3.5$), with a larger
proportion of late-type galaxies than early-type galaxies.

\clearpage

\section{Properties of Lopsidedness \label{sec:lopprops}}

\subsection{Light vs. Mass Distributions \label{sec:lightmass}}

The Fourier modes describe the two-dimensional {\em light}
distribution of galaxies.  We wish to link these modes, specifically
the lopsidedness, with a description of the two-dimensional {\em mass}
distribution as seen along the same line of sight.  If the contrast
between the bright and dim halves of a galaxy is significantly different in the
$g$-band than in the $i$-band, the lopsidedness measure may be
indicating an asymmetry in mass-to-light ratios from asymmetrically distributed  star formation and/or dust extinction.  On the other hand, similar values in
the two bands would suggest similar mass-to-light ratios and a
corresponding lopsidedness in surface mass density.  To see which scenario is
prevalent, we can look at the colors and magnitude differences between
galaxies of different star formation histories and hence mass/light ratios.  Then we can compare those differences to
those of the bright and dim halves of the lopsided galaxies.

\begin{figure}[ht]
\epsscale{1.1}
%\plottwo{/home/tar/agn/asymm/plothalfmag/meddist_avgdi_avgdgi.ps}{/home/tar/agn/asymm/plothalfmag/meddist_rpdi_rpdgi.ps}
\plottwo{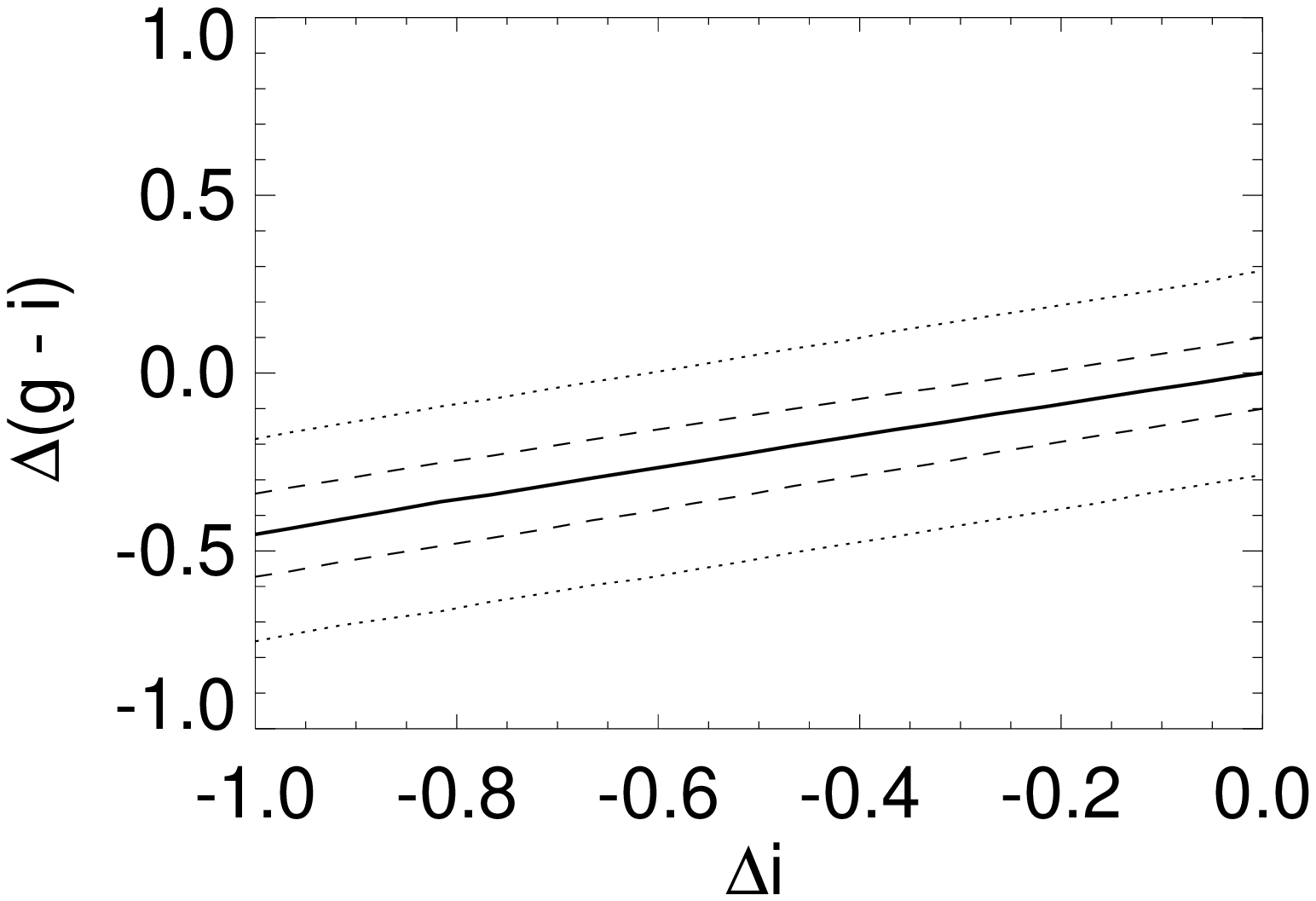}{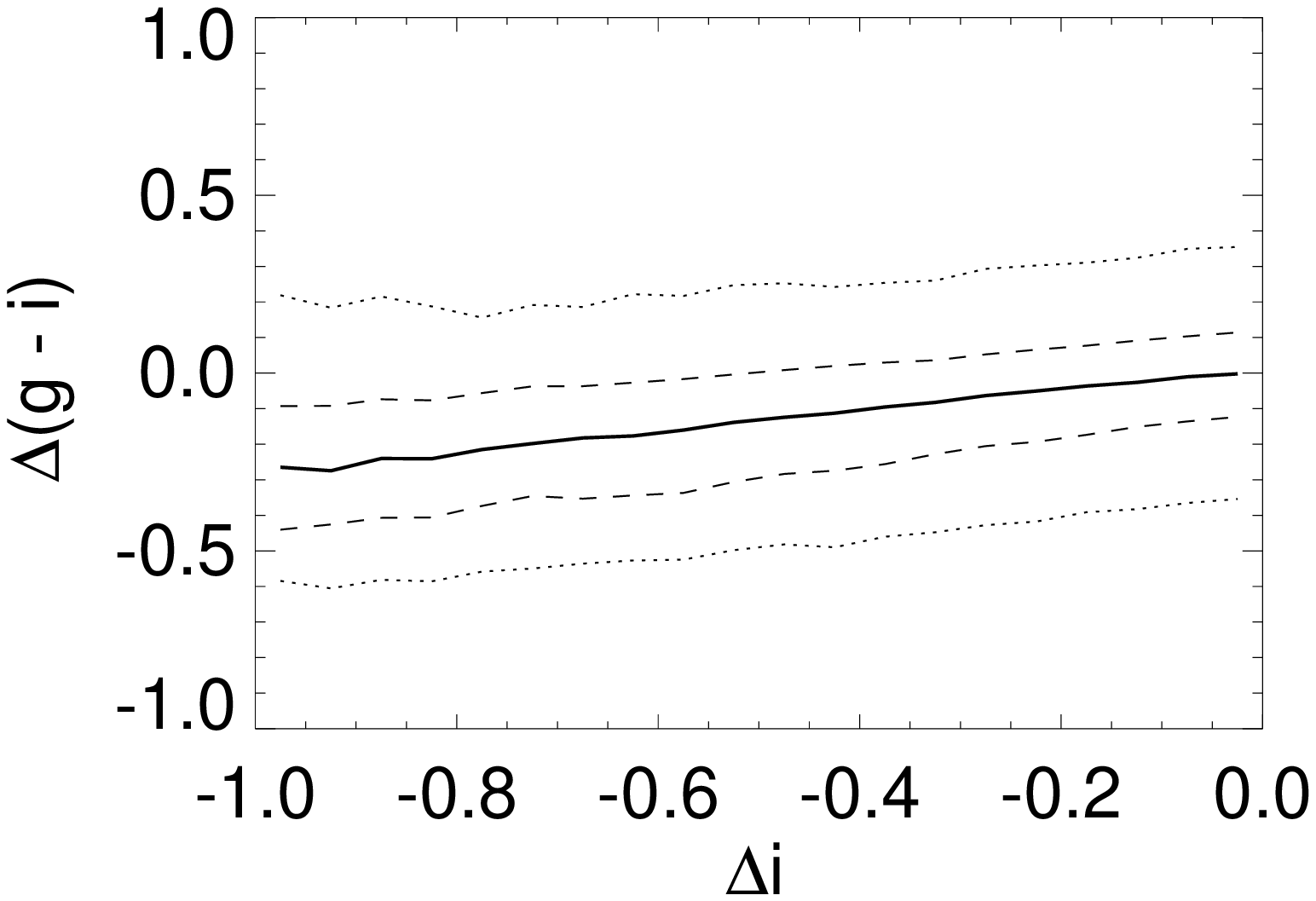}
\epsscale{0.5}
%\plotone{/home/tar/agn/asymm/plothalfmag/meddist_actdi_actdgi.ps}
\plotone{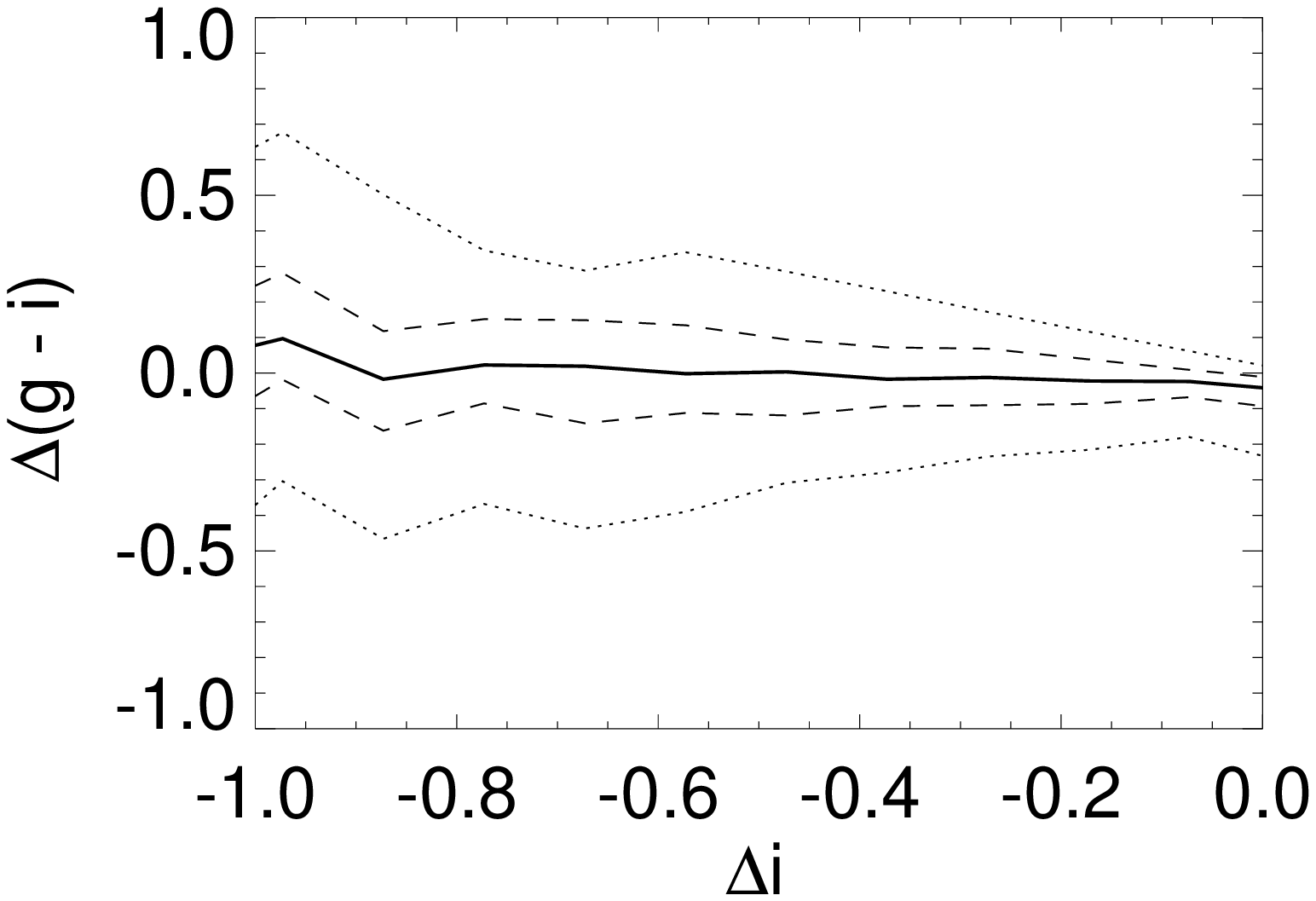}
\caption{Differences in $g - i$ color and $i$ magnitude for pairs of model stellar populations {\em (upper left)}, for pairs of observed galaxies {\em (upper right)}, and for bright and dim halves of observed galaxies {\em (lower panel)}.  The pairs of stellar populations and observed galaxies have random star formation histories and exhibit different mass-to-light ratios in the $g$ and $i$ bands.  The contrasting halves of observed galaxies instead show no correlation between color and magnitude differences, implying similar average mass-to-light ratios in each half.}
\label{fig:relcolor}

\end{figure}

We start with a set of pairs of stellar population models.  We have
taken 32000 simulations of stellar populations with identical mass and
varying star formation histories from \citet{khw+03a} and randomly
paired them to look at their relative colors.  The models were
generated from a wide range of superimposed continuous and bursty star
formation histories with varying metallicity, such that bursty and
continuously star-forming models each contribute about half of the
models.  We plot in the upper left panel of Fig.~\ref{fig:relcolor}
the distributions of the difference in color $\Delta(g-i)$ and in
magnitude $\Delta i$ within the pairs.  We find a tight relation
between the difference in brightness and color within the pair
($\Delta(g-i) = 0.45 \Delta i$ in the median).

We next perform the same comparison using 113000 pairs of low-redshift
($z < 0.06$) observed galaxies. The pairs were selected from the SDSS
DR4 galaxy sample and have been matched in redshift with $\Delta z <
0.001$ and stellar mass with $\Delta\log_{10} M_* < 0.01$.  The
relation between relative color and magnitude is shown in the upper
right panel of Fig.~\ref{fig:relcolor}.  The brighter galaxy in the
pair is the bluer galaxy, in agreement with the models. However the
relation is somewhat shallower ($\Delta(g-i) = 0.29 \Delta i$ in the
median). The difference in these relations may be due to a wider range
of SFHs, especially the more extreme, bursty SFHs, in the models than
in the observed galaxies. It may also reflect differences in the
effects of dust extinction and reddening in the models vs. the data.

We now compare these paired color-magnitude relations of the galaxies and the models to that determined from the bright and dim sides of the galaxies in our lopsided galaxy sample. Having determined the position angle that maximized the light asymmetry, we then measured the $g$, $r$, and $i$ magnitudes for each galaxy half. Stars and intervening galaxies were masked out
and excluded, as before, but for each masked
pixel here, the corresponding pixel at the same radius and $180^\circ$
away is also excluded.  Light is thus summed over the same area in
each half.  

We show in the lower panel of Fig.~\ref{fig:relcolor} the difference
in color $\Delta (g-i)$ between the contrasting halves of galaxies in
the DR4 sample as a function of their difference in $i$ magnitude. The
median color difference is zero and has little dependence on relative brightness. This is completely different from the behavior of the models or real galaxies as the mass/light ratio is varied, and suggests that the lopsidedness is not due to variations in mass/light ratio. In fact, \citet{k+07} have used SDSS galaxy spectra to show that the $(g-i)$ fiber color is an excellent proxy for the stellar mass/light ratio. The fact that there is no systematic difference in $(g-i)$ color between the brighter and dimmer sides of the lopsided galaxies then implies that the lopsided light distribution is primarily tracing a lopsided stellar mass distribution.

While there is no systematic offset in color with magnitude, we note that the spread in $\Delta (g-i)$ increases as $\Delta i$ increases. This implies that there are large-scale spatial variations in the mass/light ratio in lopsided galaxies (presumably due to enhanced star formation and/or dust extinction).

\clearpage

\begin{figure}[ht]
\epsscale{1.1}
%\plottwo{/home/tar/agn/asymm/lophists/banddist_a1.ps}{/home/tar/agn/asymm/lophists/banddist_a2.ps}
\plottwo{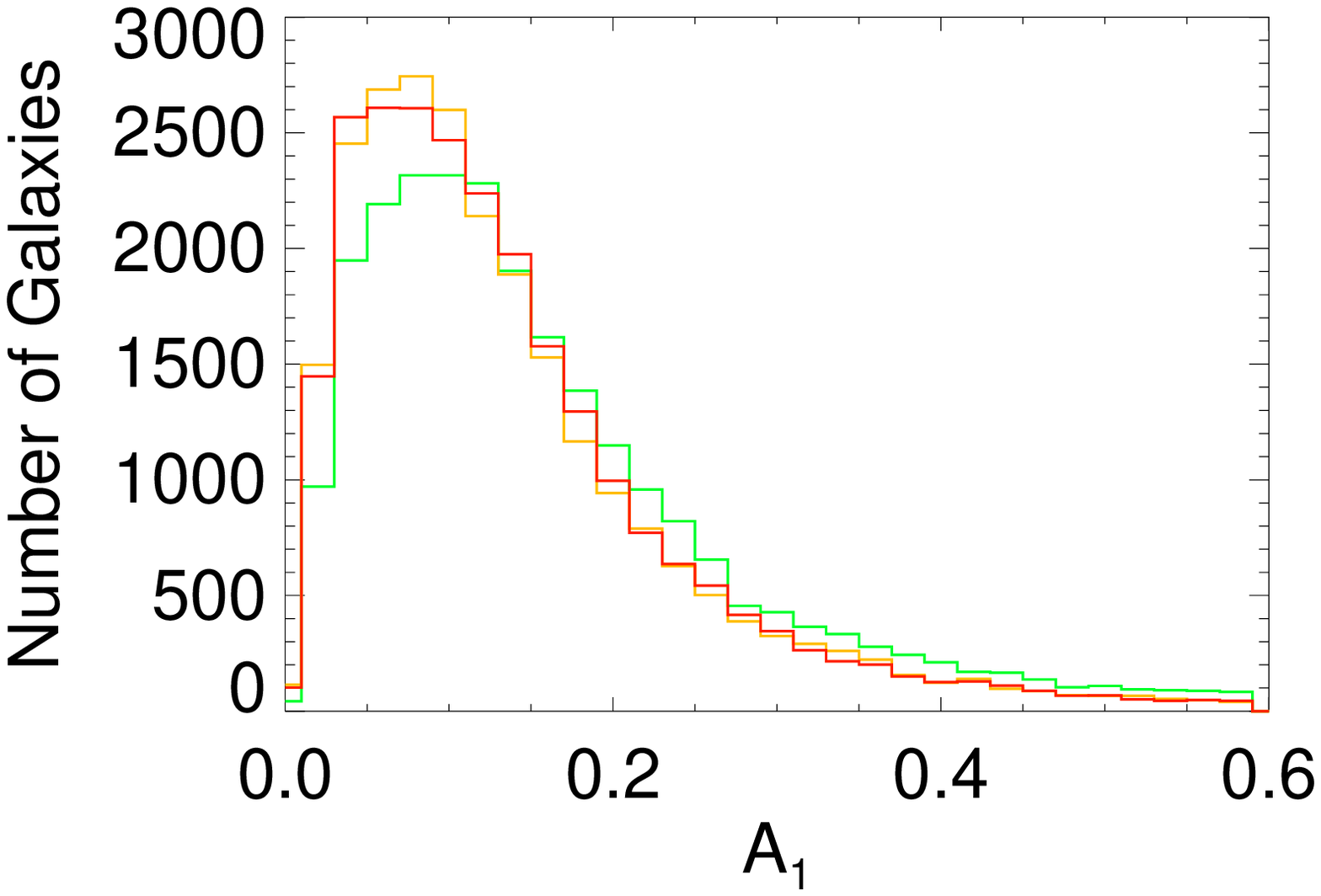}{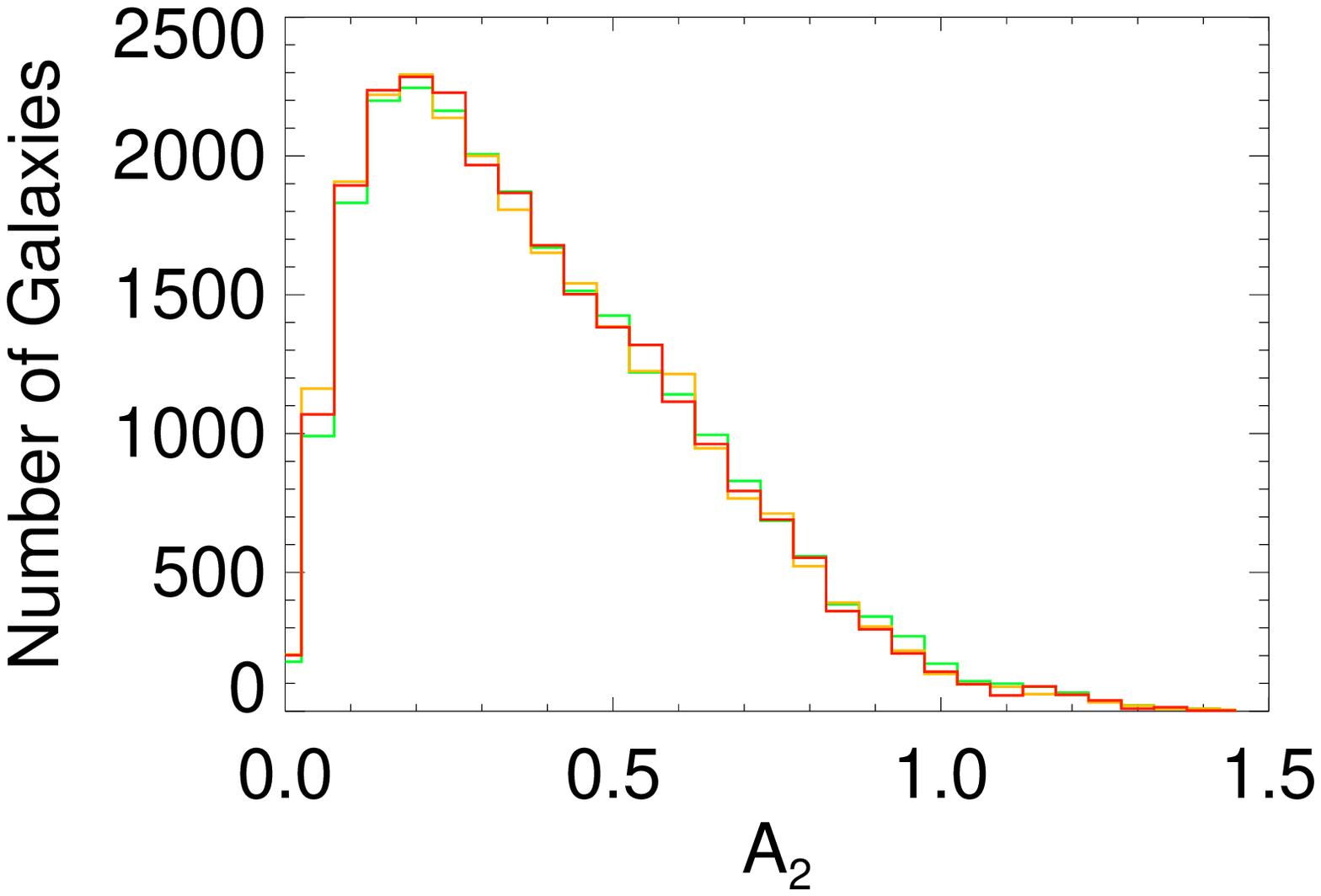}
\caption{Distributions of the first and second azimuthal Fourier modes in the $g-$ (green), $r-$ (orange), and $i-$ (red) bands between $R_{50}$ and $R_{90}$. The $r$- and $i$-band distributions are similar. The $g$-band distribution is also similar but is skewed toward slightly higher values of lopsidedness.  The lopsidedness of the light distributions is mainly tied to a lopsidedness in the mass distribution, but there is a small contribution from lopsidedness in the distribution of star formation.}
\label{fig:modesgri}
\end{figure}

As another way of addressing the importance of variations in the the mass/light ratio in causing lopsidedness, we have compared the distribution of $A_1$ in three different bands in Fig.~\ref{fig:modesgri}. The distribution measured from $g$-, $r$-, and $i$-band images are shown in green, orange, and red, respectively.  

The distributions of $A_1^r$
and $A_1^i$ are nearly identical, but the $A_1^g$ distribution is
skewed toward slightly higher values. We have already
shown that the lopsidedness of the light distribution primarily traces
the lopsidedness of the underlying mass distribution.  Here we see the
weaker, secondary effect. Newly formed stars are not uniformly
distributed, and so the lopsidedness of the light distribution
includes a small contribution from the lopsidedness of the
distribution of star formation.

\clearpage

\subsection{Radial Dependence of Lopsidedness \label{sec:raddep}}

Previous studies have shown that asymmetry in galaxies shows a clear
radial dependence.  \citet{rr98} used a sample of 54 face-on,
early-type disk galaxies and found that an increase in $A_1$
lopsidedness with radius was normal. \citet{cbj00} studied radial
profiles of both elliptical and disk galaxies in a sample of 113 from
the \citet{fgg+96} sample.  They found that asymmetry typically
increases at larger radii $r > R_{50}$ in disk galaxies but peaks at
smaller radii in elliptical and lenticular galaxies.  One would expect
our larger sample of galaxies to confirm the dependence of
lopsidedness on both radius and Hubble type.  In what follows, we use
concentration as a measure of Hubble type.

\begin{figure}[ht]
\epsscale{1.1}
%\plottwo{/home/tar/agn/asymm/lophists/prof_r_a1_lowa1.ps}{/home/tar/agn/asymm/lophists/prof_r_a1_mida1.ps}
\plottwo{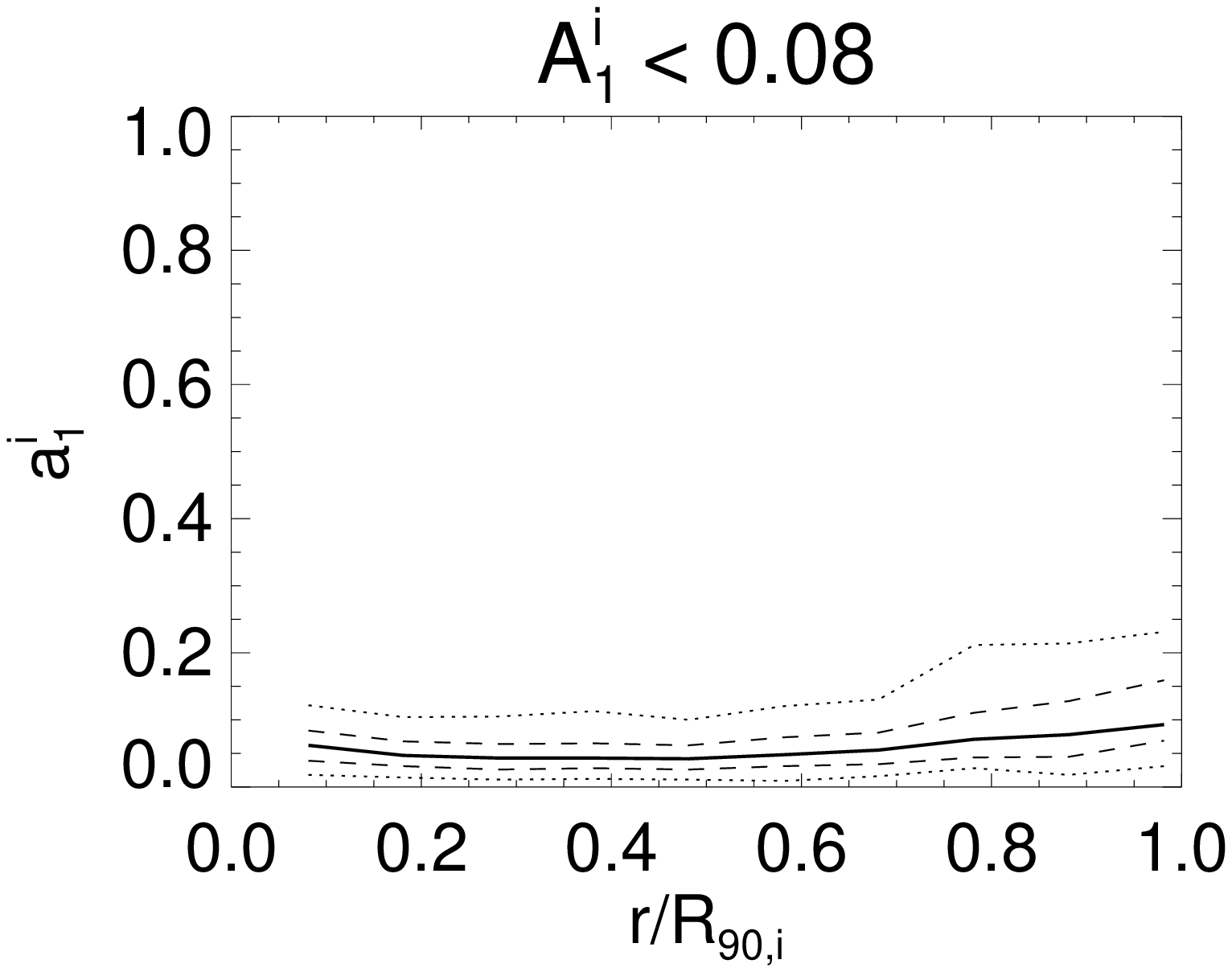}{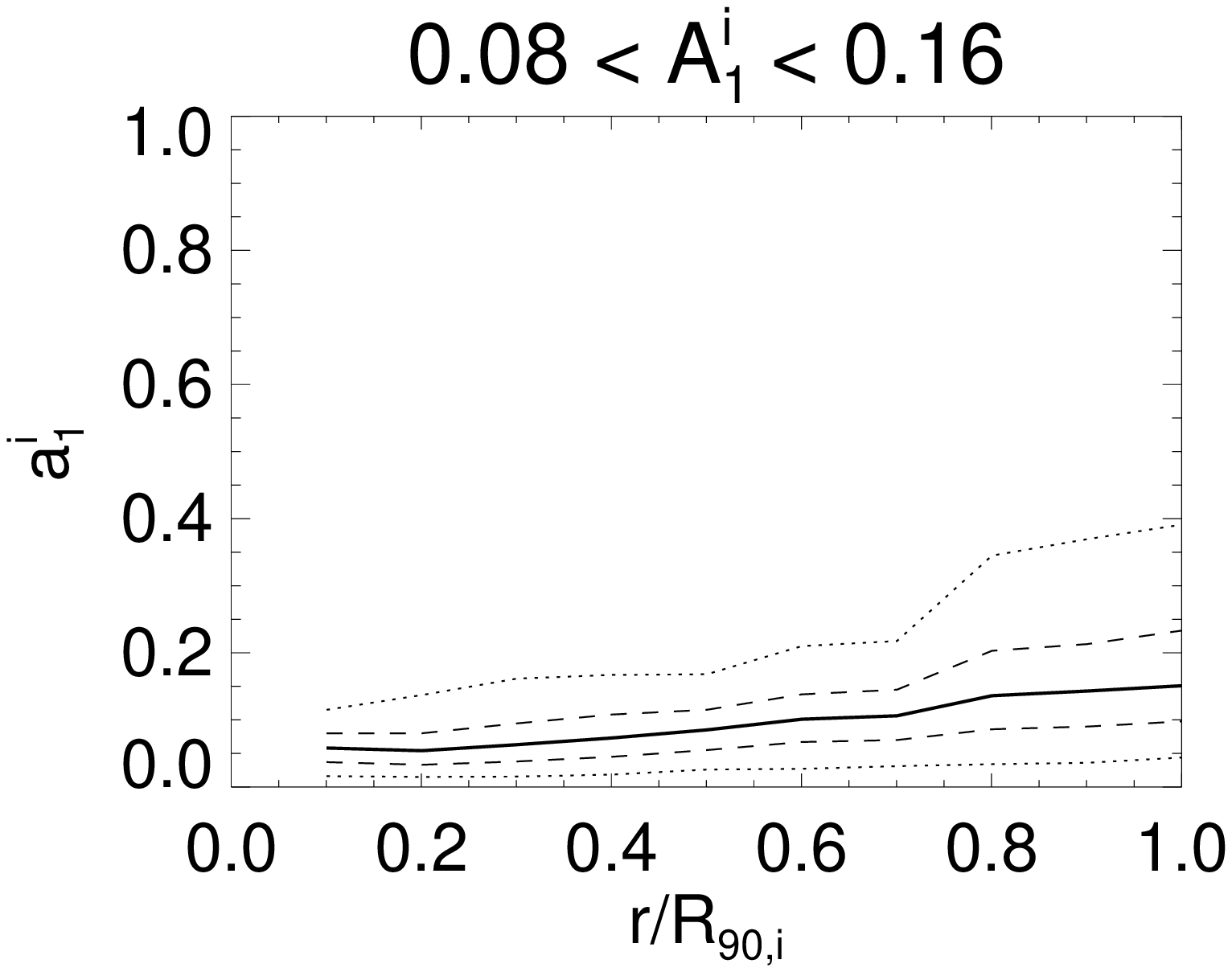}
%\plottwo{/home/tar/agn/asymm/lophists/prof_r_a1_hia1.ps}{/home/tar/agn/asymm/lophists/prof_r_a1_etg.ps}
\plottwo{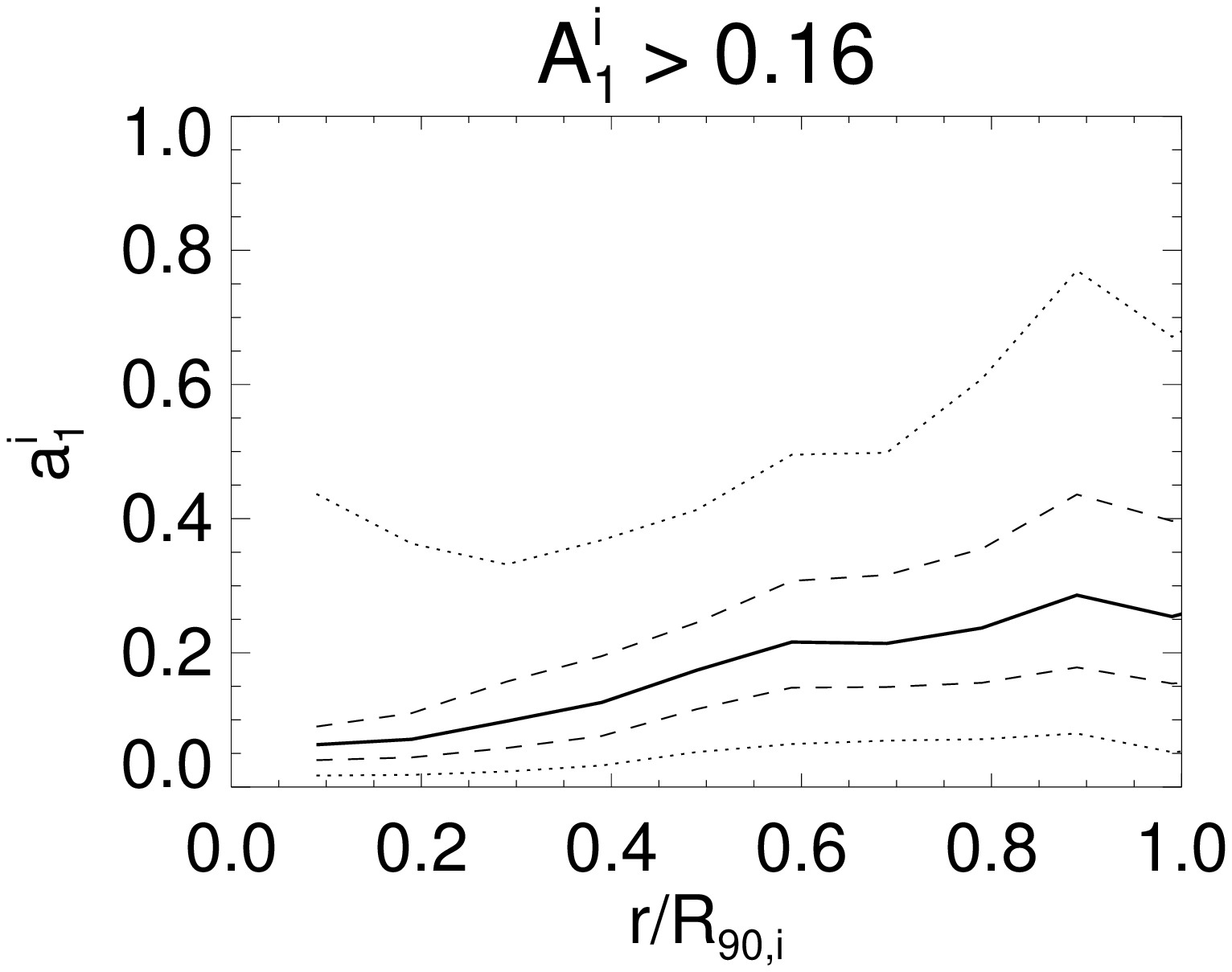}{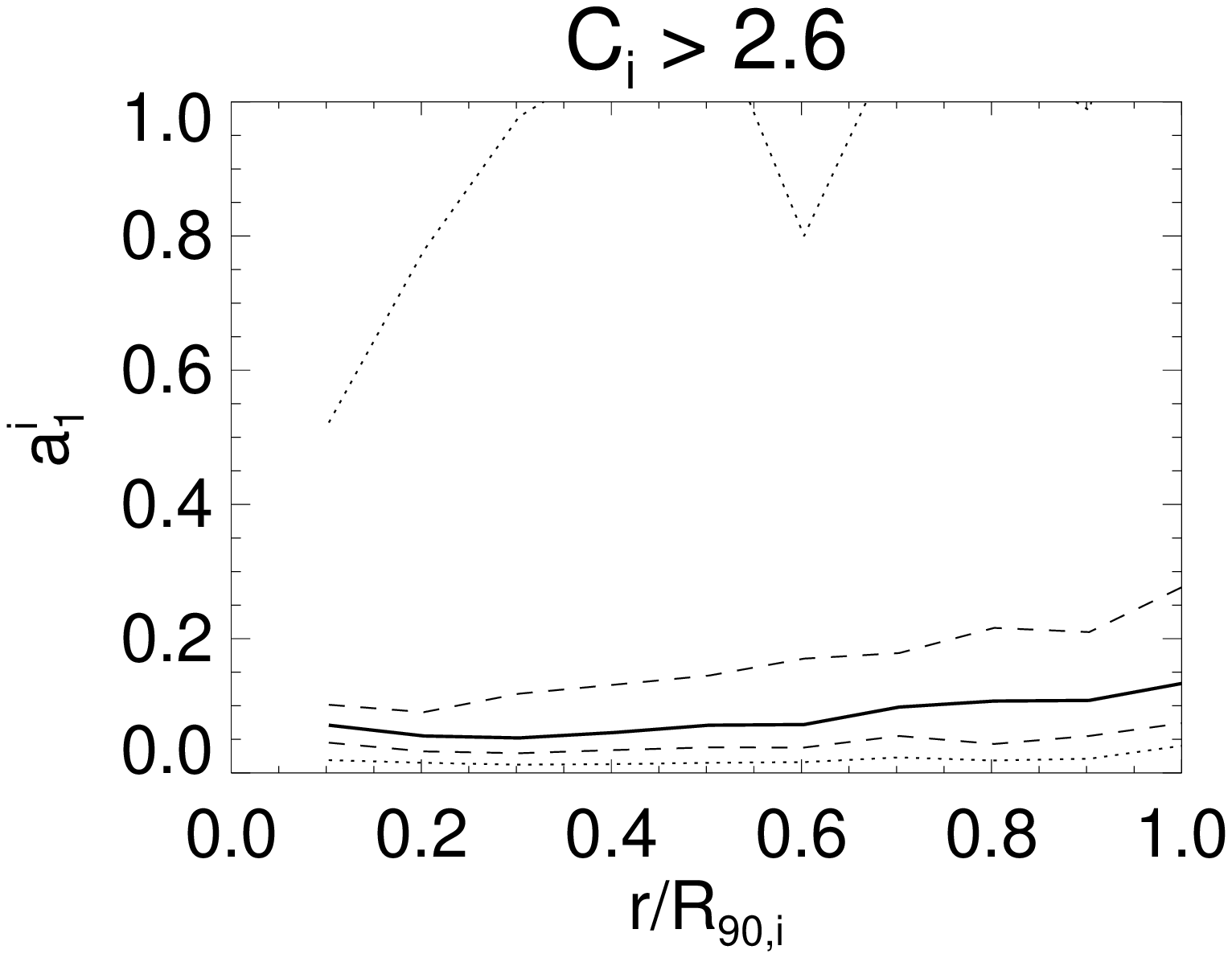}
\caption{Radial profiles of lopsidedness shown as distributions of $a_1^i$ as a function of radius.  The first three panels show distributions for symmetric, average, and lopsided late-type galaxies ($C_i < 2.6$), and the lower right panel shows the distribution for all early-type galaxies ($C_i > 2.6$).  The 5th, 25th, 50th, 75th, and 95th percentiles are shown.  Lopsidedness increases with radius in symmetric and lopsided, late-type and early-type galaxies alike.}
\label{fig:lopradprof}
\end{figure}
\clearpage

In Fig~\ref{fig:lopradprof}, we show radial profiles as the
distribution of the first Fourier mode (lowercase) $a_1^i(r)$ at radii
extending outward to $R_{90}$.  The first three panels show profiles
for late-type galaxies ($C_i < 2.6$) partitioned into bins of global
lopsidedness (capitalized $A_1^i$), and the last panel shows
early-type ($C_i > 2.6$) galaxy profile. $R_{50}$ corresponds to
roughly $0.38R_{90}$ to $0.60R_{90}$ for these late-type galaxies.  In
all these profiles, lopsidedness increases steadily at radii larger
than $\sim0.5R_{90}$.  Moderately and highly lopsided ($A_1^i > 0.08$)
late-type galaxies also show a gradual increase in lopsidedness with
radius at smaller radii.  The early-type and more symmetric late-type
galaxies show a minor decline in lopsidedness at small radii and an
increase at larger radii. The decline at small radii may be an effect
originating from minor centering errors near the galactic center,
causing a small overestimation of $a_1^i (r)$ at the smallest radii.
This effect diminishes with radius. At large radii, the increase in
$a_1^i (r)$ is tied to a real increase in light asymmetry, though
there could also be some unphysical enhancement due to the low $S/N$.

%\subsection{Catalog of Most Lopsided Galaxies}

%Create an electronic catalog of galaxies that reports their
%lopsidedness and other modes at varying radii and with other pertinent
%quantities.

\section{Structural Properties of Lopsided Galaxies \label{sec:lopsfh}}

In \S\ref{sec:cuts}, we showed that our working sample was drawn from
a full sample of $z < 0.06$ galaxies and was selected based on cuts on
several observational parameters.  The proportion of high-mass-density
and highly concentrated galaxies was reduced, but a significant number
of these galaxies was also retained.  Our sample thus allows us to study
lopsidedness of galaxies over a wide range of the basic structural properties of the galaxies, namely their concentration, stellar mass, and stellar mass density.

\begin{figure}[ht]
\epsscale{1.1}
%\plottwo{/home/tar/agn/asymm/lophists/meddist_a1i_conci.ps}{/home/tar/agn/asymm/lophists/meddist_a1i_m.ps}
\plottwo{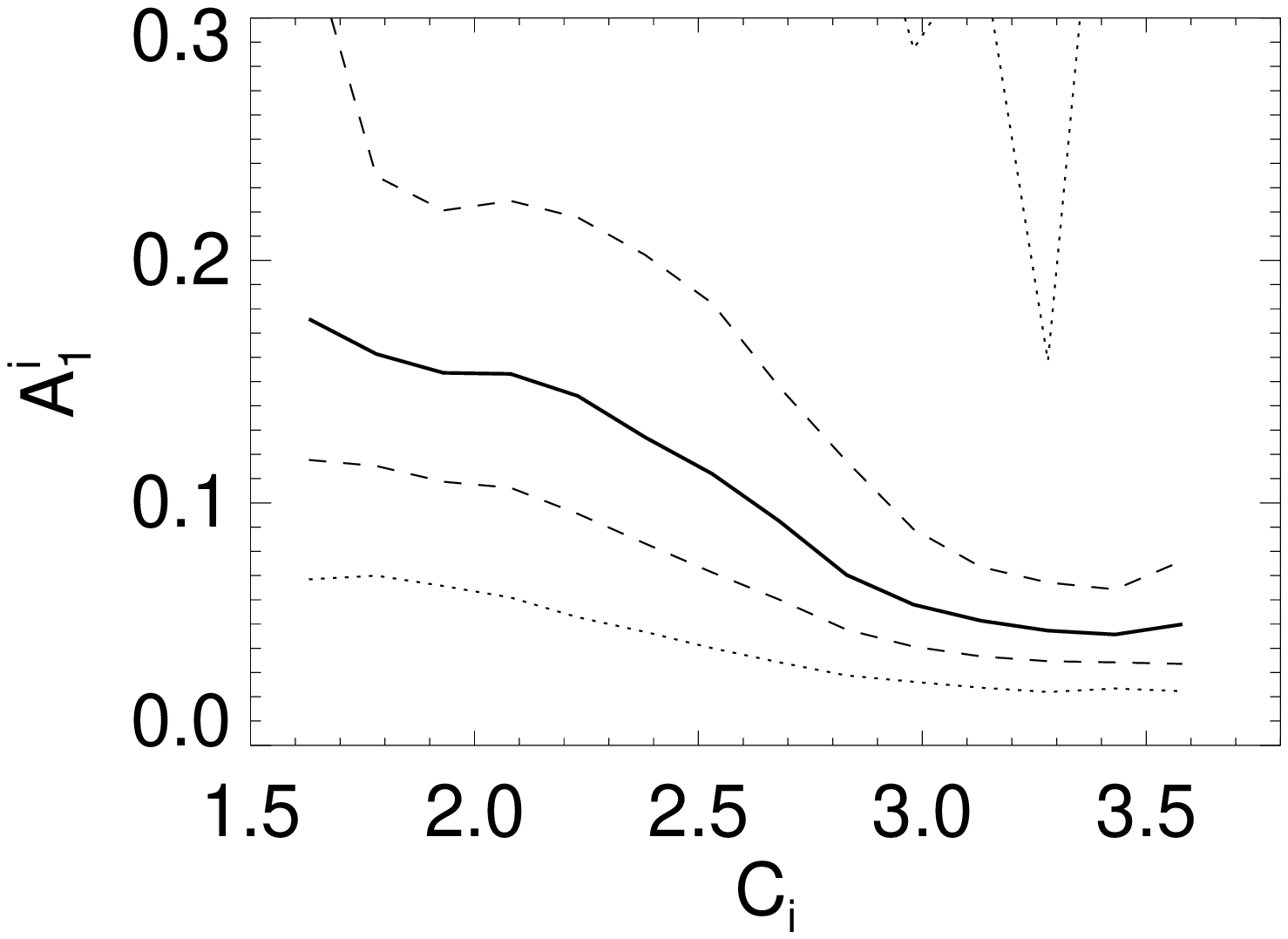}{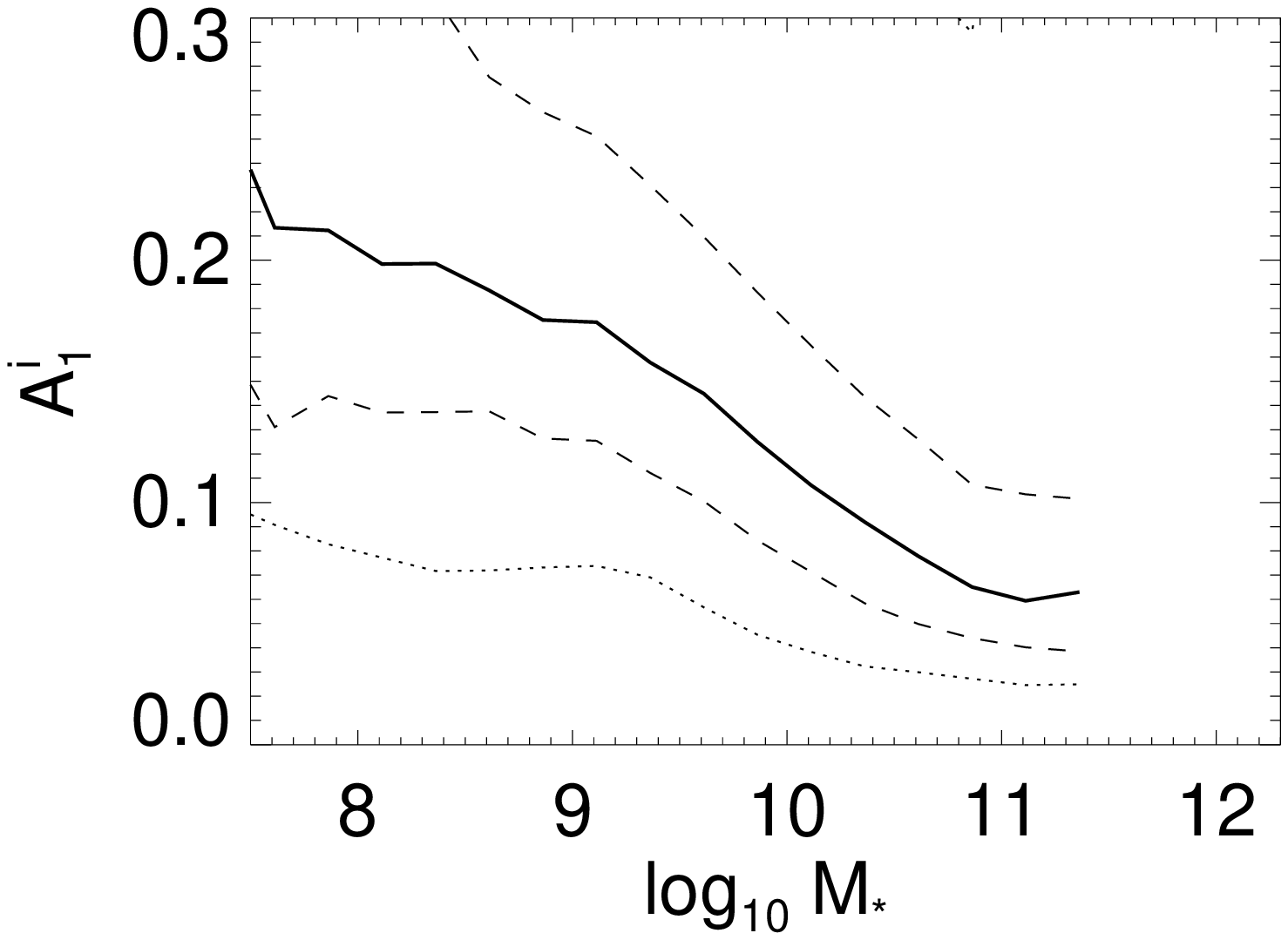}
\epsscale{0.5}
%\plotone{/home/tar/agn/asymm/lophists/meddist_a1i_mu.ps}
\plotone{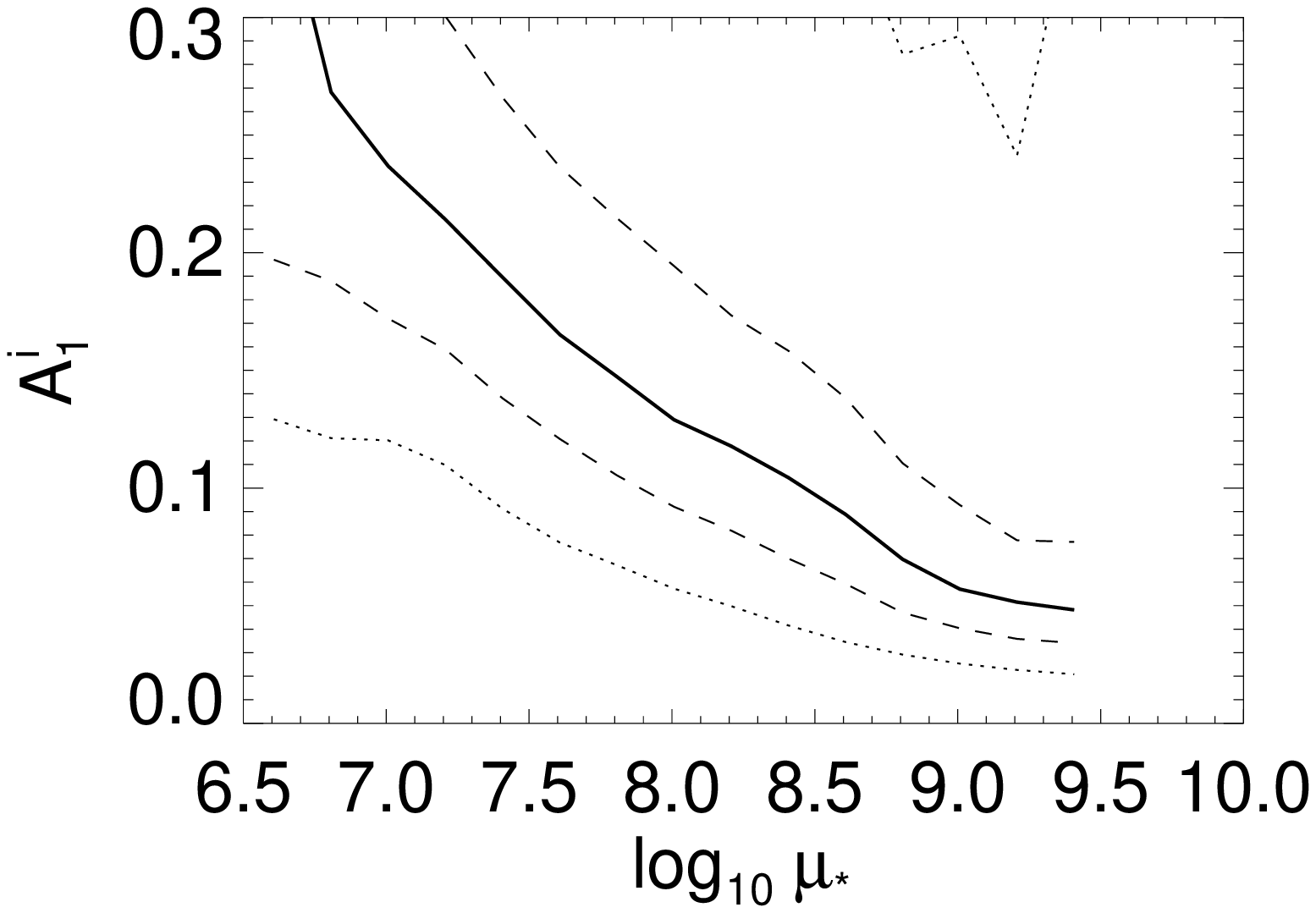}
\caption{Distributions of $i$-band lopsidedness as functions of the structural parameters $C_i$, $M_*$, and $\mu_*$.  The 5th, 25th, 50th, 75th, and 95th percentiles are shown in each panel.  Lopsided light distributions are more common in low-mass, low-mass-density, and low-concentration galaxies. Massive, dense, and concentrated galaxies tend to be considerably more symmetric.}
\label{fig:lopstruct}
\end{figure}

The concentration is a rough proxy for Hubble type, with higher values
corresponding to earlier types. The correspondence between $C_i$ and
Hubble Type has been considered by \citet{si+01} and \citet{s+01}. The
correspondence is not tight, but the value $C_i$ = 2.6 is the rough
dividing line between early- and late-type galaxies (see also
\citealt{khw+03a}. The top-left panel of Fig.~\ref{fig:lopstruct}
shows the distribution of the global $A_1^i$ as a function of
concentration. At $C_i = 2.6$, $A_1^i$ has a moderate value near 0.10.
For $C_i > 3.1$ (typical of elliptical galaxies), the lopsidedness is
typically too small to reliably measure. Below $C_i = 2.6$ the
lopsidedness rises systematically with decreasing concentration (later
Hubble types). Lopsidedness is thus commonplace in the late-type field
galaxies and the galaxies in low-density environments that dominate
our sample, in agreement with previous studies (\citealt{mv+98};
\citealt{cbj00}; \citealt{bc+05}).  However, this result might not extend to other environments \citep{aj+06,aj+07}.

The top-right panel of Fig.~\ref{fig:lopstruct} shows the distribution of lopsidedness as a function of stellar mass. \citet{khw+03b} showed that the local galaxy population is bimodal, with the transition from late type star forming galaxies to early type old galaxies occurring at a value of $M_* \sim 10^{10.5} \Msun$. We see that lopsidedness is only significant for the low mass population, and increases systematically with decreasing mass.

Finally, we show that there is a similar but even stronger trend
between lopsidedness and stellar mass density, as shown in the lower
panel of Fig.~\ref{fig:lopstruct}. \citet{khw+03b} showed that the
transition between late type star forming galaxies and early type old
galaxies occurs at a value $\mu_* \sim 10^{8.5} \Msunkpc$. We find
that lopsidedness is only commonplace among the low density
population. The lopsidedness increases very strongly with decreasing
density in this regime. Indeed at the lowest densities ($\mu_* <
10^{7.5} \MSunkpc$) typical galaxies are significantly lopsided
($A_1^i > 0.20$), and few galaxies in this density range are less
lopsided than 0.10.

\clearpage

\begin{figure}[ht]
\epsscale{1.1}
%\plottwo{/home/tar/agn/asymm/lophists/2d_m_mu_lopi.ps}{/home/tar/agn/asymm/lophists/2d_m_conci_lopi.ps}
\plottwo{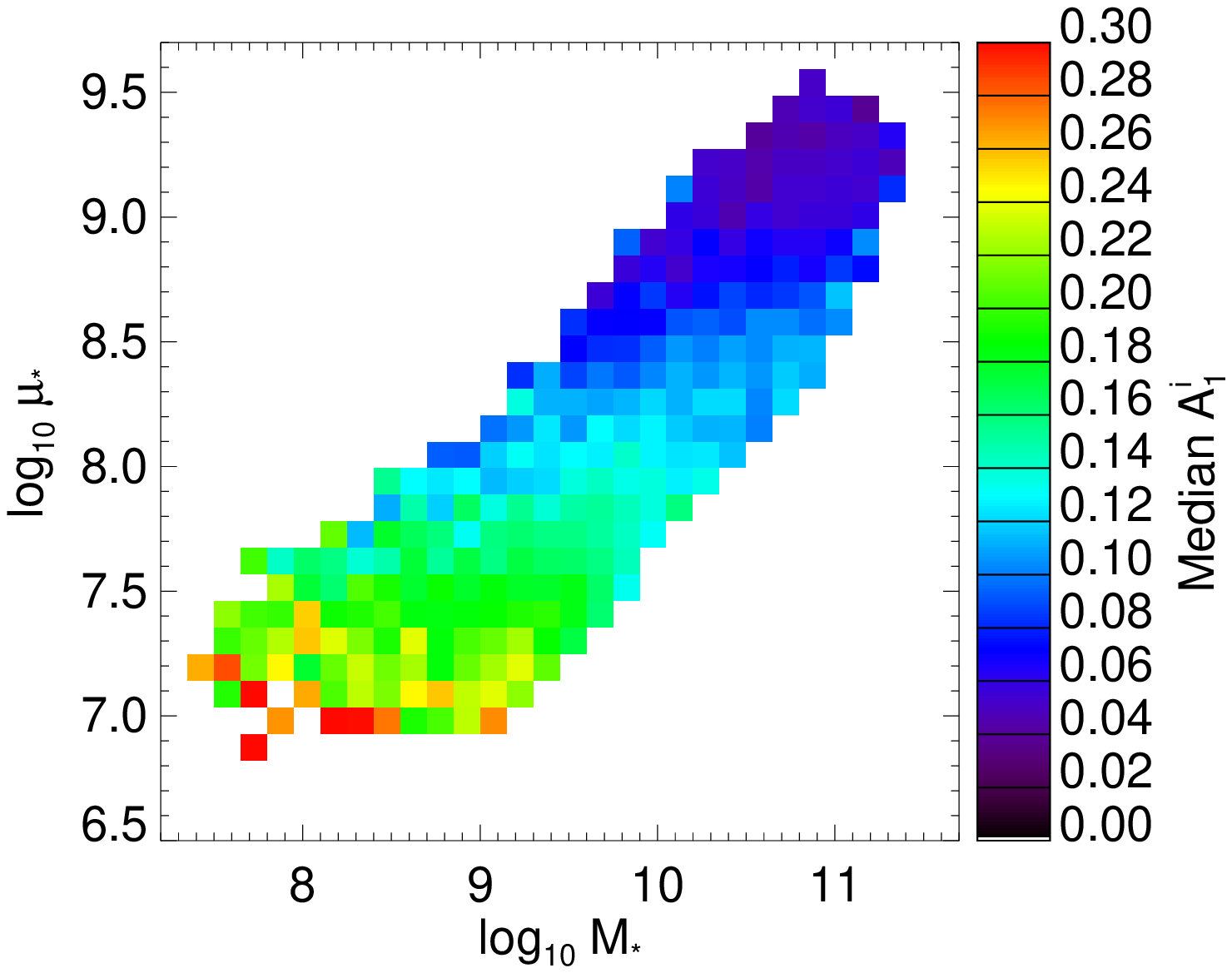}{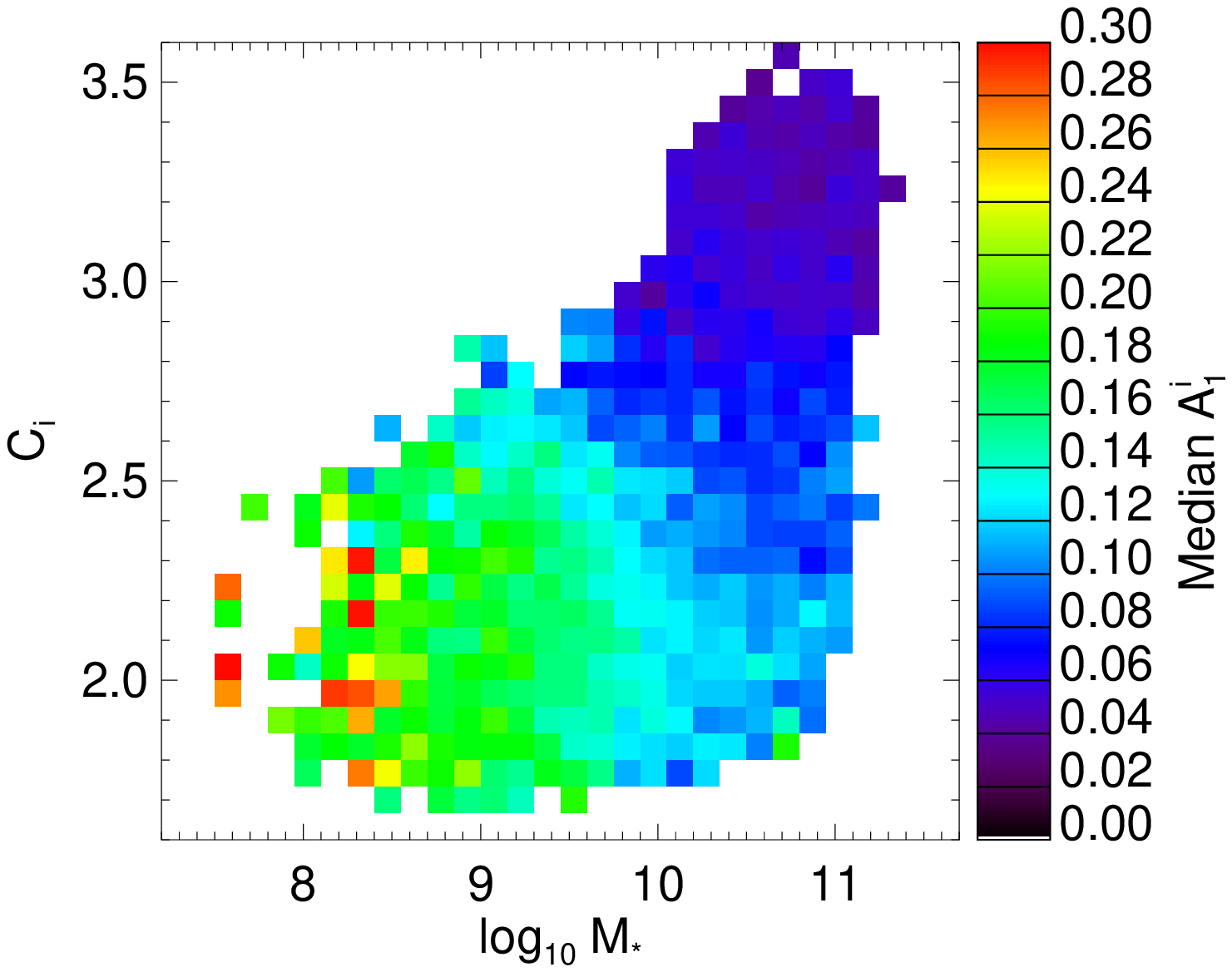}
\epsscale{0.5}
%\plotone{/home/tar/agn/asymm/lophists/2d_mu_conci_lopi.ps}
\plotone{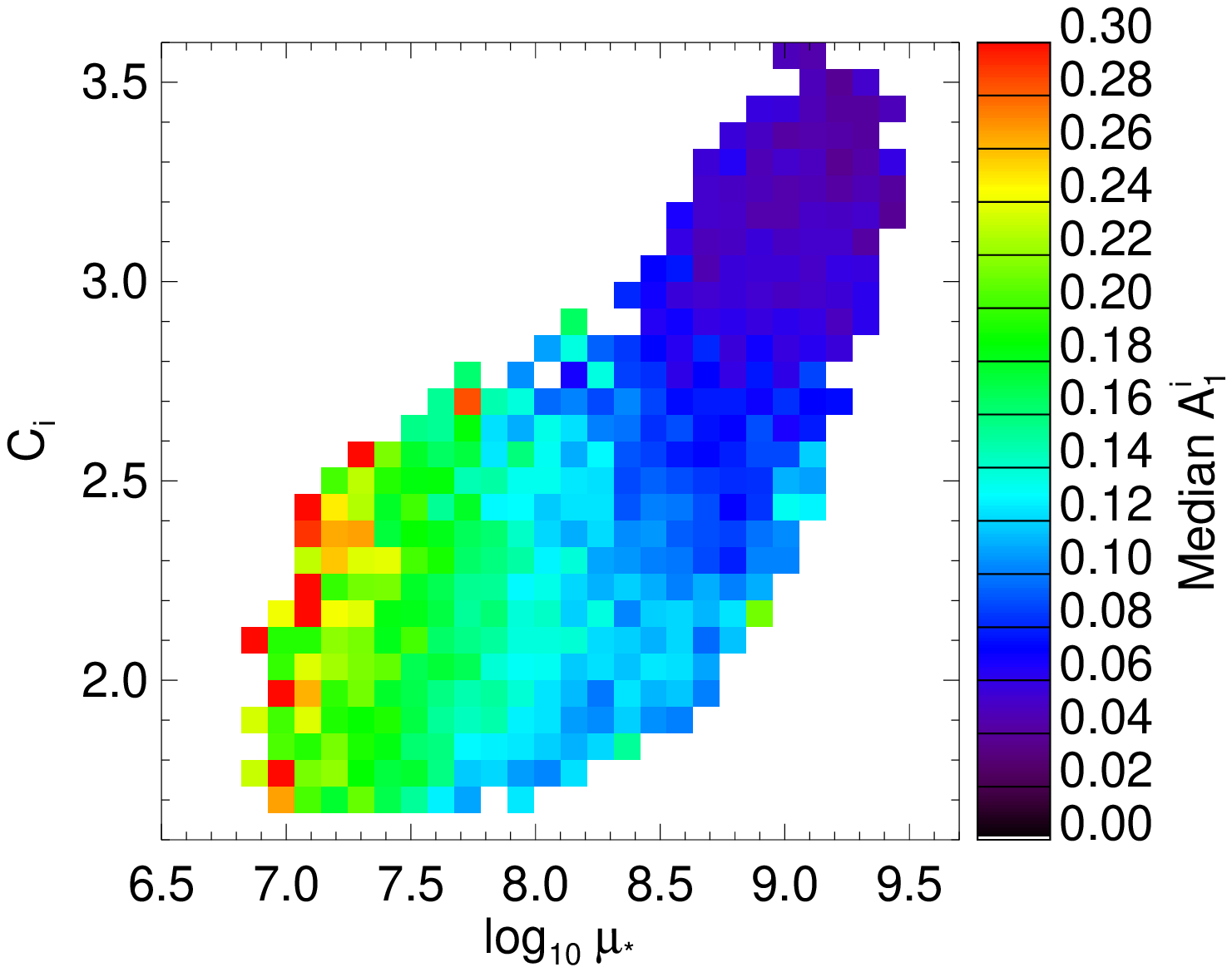}
\caption{Stellar mass-mass density and concentration-mass relationships colored by median $i-$band lopsidedness. Galaxies that are massive, dense, and concentrated tend to have symmetric light distributions, while low-mass, low-mass-density, and low-concentrated galaxies tend to be lopsided.}
\label{fig:lop2dstr}
\end{figure}

The above structural parameters of SDSS galaxies ($C, M_*, \mu*$) are
highly correlated with one another (e.g., \citealt{khw+03a}). From
the above plots, it is not immediately clear whether there is a
separate underlying physical correlation between lopsidedness and each
of the structural parameters, or whether some of the apparent
correlations are induced by mutual correlations between the
parameters. To explore this, we now turn to Fig.~\ref{fig:lop2dstr}
and compare the three structural properties, two at a time, with
lopsidedness. First, we have plotted $\mu_*$ and $C_i$ against $M_*$
in two-dimensional bins and color-coded each bin holding at least 5
galaxies by its median $A_1^i$.  We see immediately that while
lopsidedness does indeed simultaneously correlate with all these
structural parameters, the correlation with $\mu_*$ is the strongest
and most fundamental. In the plots of $\mu_*$ {\it vs.} $M_*$ and
$C_i$ we see that the lopsidedness at a given value of $\mu_*$ is
essentially independent of either $M_*$ or $C_i$. Conversely, for
given values of $M_*$ or $C_i$, there is a systematic increase in
lopsidedness with decreasing $\mu_*$. The plot of $C_i$ {\it vs.} 
$M_*$ implies that both parameters are correlated with lopsidedness
(neither is clearly the more fundamental). 

We have used linear partial correlation analysis to see which pairs of
these four structural properties show more fundamental correlations.
We calculated the linear (Pearson) correlation coefficients for each of
the six combinations of the four parameters.  We use $\log M_*$, $\log
\mu_*$, $\log A_1^i$, and $C_i$ for this analysis because the
relationships are more linear if logarithms are used for three of the
parameters.  For each pair, we also removed the dependence of the
remaining two parameters and calculated the partial correlation
coefficient.  The coefficients are listed in Table~\ref{tab:parcor}.
Lopsidedness correlates moderately with all three of the other
parameters but most strongly with stellar mass density
(corr. coeff. $= -0.56$), then concentration ($-0.47$) and mass
($-0.46$).  Once the correlations with stellar mass and concentration
are removed, the correlation between lopsidedness and mass density is
reduced in magnitude to $-0.20$. The partial correlations between
lopsidedness and the other two structural parameters are weaker
($0.00$ with $C_i$ and $-0.12$ with log $M_*$). In agreement with
Fig.~\ref{fig:lop2dstr}, we see that the most fundamentral correlation
between lopsidedness and a structural parameter is with the surface
mass density.

\begin{deluxetable}{cclr}
\tabletypesize{\small}
\tablecolumns{4}
\tablewidth{0pc}
\tablecaption{Correlation Coefficients of Structural Parameters}
\tablehead{
	\colhead{} &
	\colhead{} &
	\colhead{Dependence} &
	\colhead{(Partial)} \\
   \colhead{Par. 1} & 
	\colhead{Par. 2} & 
	\colhead{Removed} & 
	\colhead{Corr. Coeff.} }
\startdata
$\log A_1^i$ & $\log \mu_*$ & \nodata & $-$0.56 \\
%$A_1^i$ & $\mu_*$ & $M_*$ & $-$0.34 \\

%$A_1^i$ & $\mu_*$ & $C_i$ & $-$0.37 \\
$\log A_1^i$ & $\log \mu_*$ & $\log M_*$ \& $C_i$ & $-$0.20 \\

\hline
$\log A_1^i$ & $\log M_*$ & \nodata & $-$0.46 \\
%$A_1^i$ & $M_*$ & $\mu_*$ & 0.04 \\
%$A_1^i$ & $M_*$ & $C_i$ & $-$0.30 \\
$\log A_1^i$ & $\log M_*$ & $\log \mu_*$ \& $C_i$ & 0.00 \\
\hline
$\log A_1^i$ & $C_i$ & \nodata & $-0.47$ \\
%$A_1^i$ & $C_i$ & $\mu_*$ & $-0.12$ \\
%$A_1^i$ & $C_i$ & $M_*$ & $-0.30$ \\
$\log A_1^i$ & $C_i$ & $\log \mu_*$ \& $\log M_*$ & $-0.12$ \\
\hline
$\log \mu_*$ & $\log M_*$ & \nodata & $0.87$ \\
%$\mu_*$ & $M_*$ & $A_1^i$ & $0.83$ \\
%$\mu_*$ & $M_*$ & $C_i$ & $0.84$ \\
$\log \mu_*$ & $\log M_*$ & $\log A_1^i$ \& $C_i$ & $0.82$ \\
\hline
$\log \mu_*$ & $C_i$ & \nodata & $0.70$ \\
%$\mu_*$ & $C_i$ & $A_1^i$ & $0.60$ \\
%$\mu_*$ & $C_i$ & $M_*$ & $0.63$ \\
$\log \mu_*$ & $C_i$ & $\log A_1^i$ \& $\log M_*$ & $0.59$ \\
\hline
$\log M_*$ & $C_i$ & \nodata & $0.49$ \\
%$M_*$ & $C_i$ & $A_1^i$ & $0.37$ \\
%$M_*$ & $C_i$ & $\mu_*$ & $-0.32$ \\
$\log M_*$ & $C_i$ & $\log A_1^i$ \& $\log \mu_*$ & $-0.32$ \\
\enddata
\label{tab:parcor}
\end{deluxetable}

\clearpage

\section{Conclusions \& Discussion \label{sec:summary}}

We have measured large-scale galactic asymmetry for a large sample of
low-redshift ($z <$ 0.06) galaxies drawn from the Sloan Digital Sky
Survey.  Our use of lopsidedness, a radially averaged $m=1$ azimuthal
Fourier mode, has proven useful for a large fraction of the sample.
Images of a minority of galaxies in the sample have poor observational
properties that cause significant systematic errors in the lopsidedness
calculation, and these galaxies were removed from the sample via cuts
on angular size, signal-to-noise, and ellipticity/inclination.  Those
cuts removed a higher fraction of the high-mass, high-mass-density,
and high-concentration galaxies than those with low values of these
structural properties.  Nonetheless, the resulting sample is well
represented by the galaxies of the same range of structural properties
as the original sample.

We find that there are no systematic differences between the $(g-i)$
colors of the brighter and fainter sides of lopsided galaxies. This
implies that there is no systematic difference in the mass/light ratio
\citep{k+07}, and hence that the lopsided light
distributions are primarily caused by lopsided distributions in the
stellar mass. We have verified this through analysis of the
relationship between color and mass/light ratio for both model galaxy
spectral energy distributions and SDSS galaxy data. However, for our
sample the lopsidedness in the $g$-band tends to be slightly greater
than in the $r$- and $i$-bands. Thus, some of the lopsidedness in the
light does arise from the effects of star-formation and/or dust
extinction (which will more strongly affect the $g-$band light).

Lopsidedness is a structural property that depends strongly on other
structural properties.  Galaxies with progressively lower
concentration, stellar mass, or stellar mass density tend to have
progressively higher lopsidedness. We show that the strongest and most
fundamental correlation is between lopsidedness and stellar mass
density. We also find that lopsidedness increases systematically with
increasing radius, particularly for late-type galaxies.

Lopsidedness can be induced through tidal stress associated with
interactions with a companion galaxy or through accretion or minor
mergers (e.g. \citealt{zr97}; \citealt{bc+05}). Galaxies with low
density will be most affected by tidal stress, and the effects of a
tidal perturbation will last longer in such systems due to the longer
dynamical times. The same arguments pertain to the outer parts of
galaxies. Thus, the two above results make good physical
sense. Alternatively, if the dark matter halo is lopsided, its effects
on the structure of the stellar disk will be more pronounced in the
outer region and in galaxies with low mass and low density (where dark
matter is more dynamically important). The relatively large values of
lopsidedness we measure to be commonplace ($A_1 > 0.1$) appear to be
too large to be generated by internally generated dynamical processes
(e.g., \citealt{mt97}).

Our overall goal in this investigation has been to use lopsidedness as
a way of quantifying the signature of moderate or weak global
dynamical perturbations.  The next step will be to determine the
connections between such perturbations and both the on-going/recent
star formation and the growth of supermassive black holes in
galaxies. These connections can help constrain the processes and
conditions that guide the formation and evolution of the galaxies.  In
future papers we will address these questions using the present sample
of galaxies.

%are also thought to induce bursts of star formation.  Using two well
%studied SFH indicators, we have shown that lopsidedness is typically
%associated with bursty and fast star formation, and quiescent star

%formation is typical in symmetric galaxies.  We will extend this
%discussion in future work.

%Interactions are also thought to induce increased star formation rates
%in galaxies.  We have found that lopsided galaxies produce stars at
%much faster rates per unit stellar mass than symmetric galaxies.
%Lopsidedness is then another way to qualitatively describe the stellar
%population age of galaxies.  Taking lopsidedness as a signature of a
%recent interaction, the link between star formation and lopsidedness
%may be that they are symptoms of the same interactions.  We will
%extend this discussion in future work.

JB acknowledges the receipt of a FCT post-doctoral grant
BPD/14398/2003.  We would like to thank Vivienne Wild for reading a
draft of the manuscript.

Funding for the SDSS and SDSS-II has been provided by the Alfred
P. Sloan Foundation, the Participating Institutions, the National
Science Foundation, the U.S. Department of Energy, the National
Aeronautics and Space Administration, the Japanese Monbukagakusho, the
Max Planck Society, and the Higher Education Funding Council for
England. The SDSS Web Site is http://www.sdss.org/.

The SDSS is managed by the Astrophysical Research Consortium for the
Participating Institutions. The Participating Institutions are the
American Museum of Natural History, Astrophysical Institute Potsdam,
University of Basel, University of Cambridge, Case Western Reserve
University, University of Chicago, Drexel University, Fermilab, the
Institute for Advanced Study, the Japan Participation Group, Johns
Hopkins University, the Joint Institute for Nuclear Astrophysics, the
Kavli Institute for Particle Astrophysics and Cosmology, the Korean
Scientist Group, the Chinese Academy of Sciences (LAMOST), Los Alamos
National Laboratory, the Max-Planck-Institute for Astronomy (MPIA),
the Max-Planck-Institute for Astrophysics (MPA), New Mexico State
University, Ohio State University, University of Pittsburgh,
University of Portsmouth, Princeton University, the United States
Naval Observatory, and the University of Washington.

\end{document}